\documentclass[sigconf]{acmart}
\usepackage{xcolor}
\usepackage{bm}
\usepackage{multirow}
\usepackage{colortbl}
\usepackage{pifont}
\usepackage{bbding}

\definecolor{mygray}{gray}{.94}

\AtBeginDocument{%
  }

\setcopyright{none} 
\copyrightyear{2025}
\acmYear{2025}
\acmDOI{XXXXXXX.XXXXXXX}

\acmConference[MICRO 2025]{The 58th IEEE/ACM International Symposium on Microarchitecture}{October 18--22, 2025}{Seoul, Korea}

\acmISBN{978-X-XXXX-XXXX-X/XX/XX}

\begin{document}

\title{MCBP: A Memory-Compute Efficient LLM Inference Accelerator Leveraging Bit-Slice-enabled Sparsity and Repetitiveness}

\author{Huizheng Wang}
\affiliation{%
  \institution{Tsinghua University}
  \institution{School of Integrated Circuits}
  \city{Beijing}
  \country{China}}
  \email{wanghz22@mails.tsinghua.edu.cn}

\author{Zichuan Wang}
\affiliation{%
  \institution{Tsinghua University}
  \institution{School of Integrated Circuits}
  \city{Beijing}
  \country{China}}
  \email{wang.zichuan@foxmail.com}

\author{Zhiheng Yue}
\affiliation{%
  \institution{Tsinghua University}
  \institution{School of Integrated Circuits}
  \city{Beijing}
  \country{China}}
  \email{yuezh20@mails.tsinghua.edu.cn}

\author{Yousheng Long}
\affiliation{%
  \institution{Tsinghua University}
  \institution{School of Integrated Circuits}
  \city{Beijing}
  \country{China}}
  \email{longys21@mails.tsinghua.edu.cn}
  
\author{Taiquan Wei}
\affiliation{%
  \institution{Tsinghua University}
  \institution{School of Integrated Circuits}
  \city{Beijing}
  \country{China}}
  \email{weitq24@mails.tsinghua.edu.cn}

\author{Jianxun Yang }
\affiliation{%
  \institution{Tsinghua University}
  \institution{School of Integrated Circuits}
  \city{Beijing}
  \country{China}}
  \email{jianxunyang@hotmail.com}

\author{Yang Wang}
\affiliation{%
  \institution{Tsinghua University}
  \institution{School of Integrated Circuits}
  \city{Beijing}
  \country{China}}
  \email{wangyang_imec@mail.tsinghua.edu.cn}

\author{Chao Li }
\affiliation{%
  \institution{Shanghai Jiao Tong University}
  \institution{Department of Computer Science and Engineering}
  \city{Shanghai}
  \country{China}}
  \email{lichao@cs.sjtu.edu.cn}

\author{Shaojun Wei}
\affiliation{%
  \institution{Tsinghua University}
  \institution{School of Integrated Circuits}
  \city{Beijing}
  \country{China}}
  \email{wsj@tsinghua.edu.cn}

\author{Yang Hu}
\authornote{Corresponding author}
\affiliation{%
  \institution{Tsinghua University}
  \institution{School of Integrated Circuits}
  \city{Beijing}
  \country{China}}
  \email{hu_yang@tsinghua.edu.cn}

\author{Shouyi Yin}
\affiliation{%
  \institution{Tsinghua University}
  \institution{School of Integrated Circuits}
  \institution{Shanghai Artificial Intelligence Laboratory}
  \city{Beijing}
  \country{China}}
  \email{yinsy@tsinghua.edu.cn}
  




\begin{abstract}
Large language models (LLMs) face significant inference latency due to inefficiencies in GEMM operations, weight access, and KV cache access, especially in real-time scenarios. This highlights the need for a versatile compute-memory efficient accelerator. Unfortunately, existing Transformer accelerators struggle to address both aspects simultaneously, as they focus on value-level processing, missing fine-grained opportunities to optimize computation and memory collaboratively. This paper introduces MCBP, a bit-grained compute-memory efficient algorithm-hardware co-design that leverages bit-slice (BS) enabled repetitiveness and sparsity to accelerate LLM inference. MCBP features three key innovations: 1) BS-repetitiveness-enabled computation reduction (BRCR), which eliminates redundant GEMM computations via leveraging redundancy hidden among BS vectors; 2) BS-sparsity-enabled two-state coding (BSTC), which reduces weight access via exploiting significant sparsity in high-order bit-slice weight; 3) Bit-grained progressive prediction (BGPP), which reduces KV cache access by leveraging early-termination-based bit-grained prediction. These techniques, supported by custom accelerator designs, effectively alleviate the burden in GEMM, weight access, and KV cache access. Extensive experiments on 26 benchmarks show that MCBP achieves $9.43\times$ speed up and $31.1\times$ higher energy efficiency than Nvidia A100 GPU. Compared to SOTA Transformer accelerators, MCBP achieves $35\times$, $5.2\times$ and $3.2\times$ energy saving than Spatten, FACT and SOFA, respectively.
\end{abstract}


\maketitle

\section{Introduction}\label{sec:Intro}
Large language models (LLMs) are transforming AI with impressive capabilities across tasks, like code generation \cite{chen2021evaluating,roziere2023code,nam2024using}, chatbots \cite{chiang2023vicuna,taori2023stanford}. As many LLM-based services rely on real-time interactions \cite{kazemitabaar2024codeaid,patel2024splitwise,lin2024infinite}, inference latency has emerged as a critical performance metric.

LLM inference \cite{spector2023accelerating} comprises two stages: prefill and decoding. In the prefill stage, the model processes all input tokens (i.e., prompt) in parallel to generate the first token, while storing intermediate Key/Value (KV) tensors, known as the \emph{KV Cache}. In the subsequent decoding stage, the model generates tokens autoregressively, each iteration requiring access to the full model weights and KV cache. 

\begin{figure}[t]
\centering
\includegraphics[width=\linewidth]{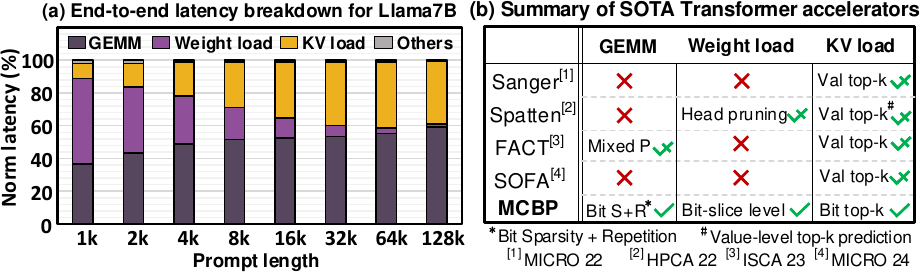}\vspace{-2mm}
\caption{(a) Key bottleneck breakdown of end-to-end latency for Llama7B (Batch=4) on NVIDIA A100 GPU with TensorRT-LLM. (b) Summary of current Transformer accelerators.}\vspace{-4mm}
\label{fig:Intro}
\end{figure}

However, the differing processing characteristics of the prefill and decoding stages, each presenting distinct resource intensities, make end-to-end inference optimization more challenging.
Fig. \ref{fig:Intro} (a) shows an end-to-end latency breakdown for LLaMA-7B under varying prompt lengths, including both the prefill and decoding stages. In this setting, the decoding is fixed at 16 tokens. The major latency contributors are categorized into GEMM computation, weight loading, KV cache loading, and others. The results indicate that all three factors significantly impact end-to-end latency across different prompt conditions. For short prompts (e.g., 1k tokens), weight loading during the decoding stage dominates, accounting for $52.4\%$ of the latency. As prompt length increases, GEMM computation in prefill stage and KV cache loading during decoding emerge as the primary bottlenecks. \textbf{These trends highlight the need for joint optimization of GEMM computation, weight access, and KV cache access to enhance end-to-end inference efficiency.}

As summarized in Fig.\ref{fig:Intro} (b), though numerous Transformer accelerators have been proposed \cite{ham20203,ham2021elsa,lu2021sanger,qu2022dota,yang2022dtatrans,zhou2022energon,wang2021spatten,qin2024mecla,wang2024sofa,qin2023fact}, most of them focus on leveraging token sparsity to mitigate the quadratic complexity of attention, which becomes a bottleneck for long inputs. Although this also partially reduces the KV load, their value-level top-$k$ prediction involves redundant computation and memory overhead, leading to inefficiency. In addition, while FACT \cite{qin2023fact} and Spatten \cite{wang2021spatten} can partially mitigate the GEMM and weight load bottlenecks via mixed-precision computation and head pruning, respectively, they lack a holistic optimization strategy that addresses all performance bottlenecks. \textbf{These limitations motivate a specialized LLM accelerator capable of jointly optimizing GEMM computation, weight access, and KV cache access.}

\begin{figure}[t]
\centering
\includegraphics[width=\linewidth]{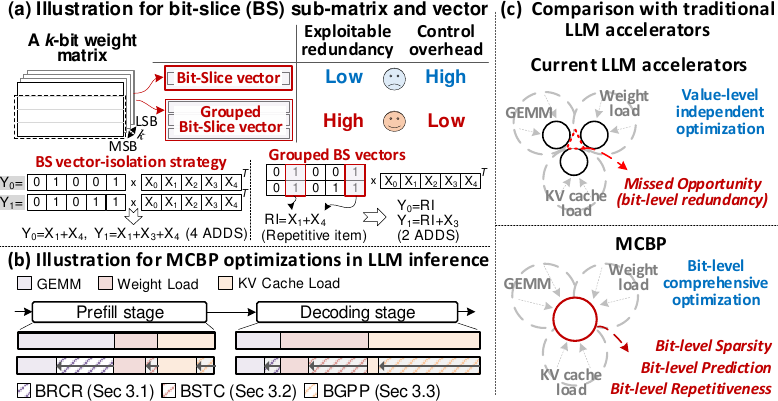}\vspace{-2mm}
\caption{Comparison between MCBP and existing works.}
\label{fig:overall}\vspace{-4mm}
\end{figure}

We conduct an in-depth analysis of the root causes behind computation and memory inefficiencies in LLM inference, identifying that these challenges can be effectively addressed via a co-designed bit-level data storage and computation scheme. As illustrated in Fig. \ref{fig:overall} (c), this work takes the first step toward addressing all major LLM bottlenecks through a unified bit-level optimization strategy.

As depicted in Fig. \ref{fig:overall} (a), we introduce the "grouped bit-slice (BS)" effect, wherein redundancy among BS vectors can be maximally exploited to reduce computation complexity while minimizing extra overhead. Throughout this paper, for an INT-quantized $k$-bit vector, it can be decomposed into $k$ individual $1$-bit vectors, and referred to as bit-slice vectors for clarity. Assuming that two bit slices of weight vectors are multiplied with a set of INT8 vectors $\mathbf{X}$, resulting in outputs $Y_0$ and $Y_1$. As exemplified in Fig. \ref{fig:overall} (a), computing $Y_0$ and $Y_1$with a naïve BS-vector-isolation strategy requires 4 additions, whereas the group BS approach needs at most 2 ADDs, by leveraging the intrinsic repetitiveness among BS vectors.

However, it is non-trivial to harness the newly identified redundancy and sparsity at bit level. The optimal granularity for bit-level processing and data compression must be carefully determined to avoid diminishing returns from control overhead due to overly fine-grained processing. Specifically, we elaborate on three key opportunities and challenges that arise from adopting bit-level processing and data compression:

a) \textbf{Unexploited redundancy among BS vectors}. Existing designs lack an effective method to exploit redundancy across BS vectors without incurring significant bit-level control overhead. While some prior works \cite{hegde2018ucnn,sharma2018bit,wang2021efficient,wang2018low} have explored leveraging redundancy across convolution channels in CNNs to accelerate computation, such techniques are not applicable to LLMs. This is due to: (1) the relatively small number of channels in CNNs, which eliminates the need for fine-grained granularity control; and (2) the small convolution kernel sizes, which incur small matching overhead for repetitive items. By contrast, the huge matrix sizes in LLMs make it challenging to efficiently identify redundancy across BS vectors in hardware, and highlight the need for a carefully selected grouping granularity to balance efficiency and overhead.

b) \textbf{Untapped sparsity resided in BS weight matrix}, due to the mismatch between value-level compression and inherent BS level sparsity. This limitation stems from the conventional value-centric memory storage paradigm, which inherently favors value-level compression techniques \cite{han2015deep,zhang2016cambricon,han2016eie}. While such methods are straightforward and widely adopted, they hinder the full exploitation of the fine-grained sparsity present at the bit-slice level. This underscores the need for a bit-dimensional compression strategy, with consistent data organization across the memory hierarchy.

c) \textbf{Inefficient Top-k prediction mechanism}, due to redundant KV cache access. 
The widely used top-$k$ mechanism in LLMs alleviates attention complexity by speculating attention sparsity and avoiding trivial token computation \cite{wang2021spatten,wang2024sofa,qin2023fact,yang2022dtatrans}. However, current value-based top-$k$ prediction is inefficient, leading to redundant IO traffic, which in turn makes the prediction itself become a bottleneck in latency. A finer-grained top-$k$ mechanism is needed to reduce I/O overhead while maintaining sparsity effectiveness.

To this end, we propose an algorithm-hardware co-design for LLM inference optimization, named MCBP. It features three key designs that correlate to three challenges, as shown in Fig. \ref{fig:overall} (b).  

1) We propose a BS-repetitiveness-enabled computation reduction (BRCR) strategy for accelerating GEMM. It identifies and reuses repetitive computations between multiple weight-BS vectors by grouping them at an appropriate granularity. This eliminates repetitive operations among grouped vectors while amortizing bit-level control overhead across them.

2) We propose a BS-sparsity-enabled two-state coding (BSTC) for weight de/compression. It strategically employs bit-slice independent encoding to exploit the significant sparsity commonly hidden in high-order bit slices. Meanwhile, we perform a joint exploration of BSTC granularity and BRCR granularity, identifying the optimal granularity configuration for seamless weight decompression and computation that maximizes overall system benefits. 

3) We design a bit-grained progressive prediction (BGPP) mechanism, to reduce unnecessary KV cache traffic during the attention sparsity prediction stage. BGPP employs a progressive bit-level filter to incrementally eliminate trivial Keys in each prediction round, enabling early termination to avoid redundant computation and memory access associated with them.

To support the above optimization mechanisms effectively, we design a dedicated accelerator named MCBP: 1) For BRCR, it employs a Content Addressable Memory (CAM) to accelerate the identification for repetitive computations, thus significantly reducing merging latency for repetitive operations. 2) For BSTC, lightweight and customized encoders/decoders are designed to enhance data compression and decompression efficiency. Additionally, it reformulates the data layout in memory to facilitate seamless bit-prioritized computation, thereby reducing data reorder overhead. 3) For BGPP, a bit-grained adaptive threshold-aware clock-gated prediction module is designed to achieve low-power attention sparsity prediction. The MCBP accelerator achieves an average energy efficiency of $22740$ GOPS/W, which is $31.1\times$, $35\times$, $5.2\times$ and $3.2\times$ higher than A100 GPU, SOTA accelerator Spatten, FACT and SOFA, respectively.

\section{Background and Motivation}\label{sec:Background}
\subsection{Large Language Models (LLMs)}\label{subsec:LLM}
LLMs \cite{anil2023palm,chang2024survey,talebi2023beyond} are based on Transformer architectures \cite{vaswani2017attention}. Initially, a length-$S$ sequence is projected into three spaces, termed Query (Q), Key (K) and Value (V), respectively. Next, Q and K are multiplied to generate an attention matrix $\mathbf{A}$ with $\mathbb{R}^{S\times S}$, which represents the correlation of each token pair. The attention matrix is then passed through a softmax operation and multiplied with V activation, resulting in a matrix $\mathbf{O}\in \mathbb{R}^{S\times H}$, where $H$ denotes hidden dimension. Finally, a feed-forward network generates the output results. 

\textbf{LLM Integer Quantization}. Quantization reduces LLMs' compute and memory costs. Early work like Q8-BERT \cite{zafrir2019q8bert} achieves INT8 weight quantization with minimal accuracy loss. In 2022, LLM.int8 \cite{dettmers2022gpt3} scaled INT8 quantization to 175B models with few INT16 outliers. SmoothQuant \cite{xiao2023smoothquant} later enabled lossless 8-bit weight and activation quantization for LLMs with up to 530B. In 2024, Atom \cite{zhao2024atom} implements 8-bit KV cache quantization. Quantization has become a prominent trend for deploying LLMs, supported by frameworks like TensorRT \cite{Nvidia2023}. Thus, optimizing compute and memory access for integer-quantized LLMs is an increasingly critical topic.

\subsection{Attention Sparsity and Top-$k$ Prediction}\label{subsec:Sparsity_in_Atten}
The standard attention mechanism in LLMs captures global context correlation via dense attention matrices. However, weak correlations between tokens produce many small attention scores, which are further suppressed by the softmax operation, further pushing them toward zero. This creates opportunities for \emph{attention sparsity}.

\begin{figure}[t]
\centering
\includegraphics[width=\linewidth]{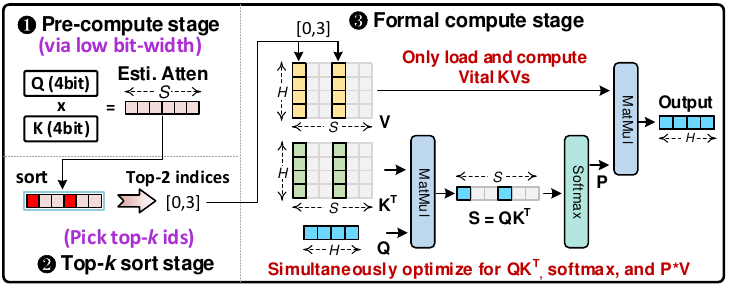}\vspace{-3mm}
\caption{Top-$k$ sparsity prediction for attention acceleration.}
\label{fig:DS}\vspace{-3mm}
\end{figure}

To exploit the \emph{attention sparsity} for computation acceleration, the top-$k$ prediction mechanism has been proposed \cite{ham20203}. Fig. \ref{fig:DS} illustrates its workflow via a $1\times S$ attention example. Typically, it consists of three stages. Firstly, a \emph{Pre-compute stage} estimates the attention matrix with a low-overhead paradigm (e.g., 4\,bit MSB). The \emph{Top-k sort stage} then selects the indices of the top-$k$ highest-scoring Keys for each Query. For example, in the estimated attention in Fig. \ref{fig:DS}, Keys [0, 3] are identified as top-2 candidates for the current Query. Finally, the indices [0, 3] are transferred to \emph{Formal compute stage}, which performs full-precision $\mathbf{QK}^T$ (8bit), softmax (FP16) and $\mathbf{PV}$ (8bit), using only these selected Keys and Values (i.e., [0, 3]). The top-$k$ mechanism has been widely adopted in recent accelerators \mbox{\cite{wang2021spatten,wang2024sofa,qin2023fact,yang2022dtatrans}} to improve attention efficiency.

\subsection{Opportunity and Observation at Bit-level}\label{subsec:Opportunity}
\begin{figure}[t]
\centering
\includegraphics[width=0.99\linewidth]{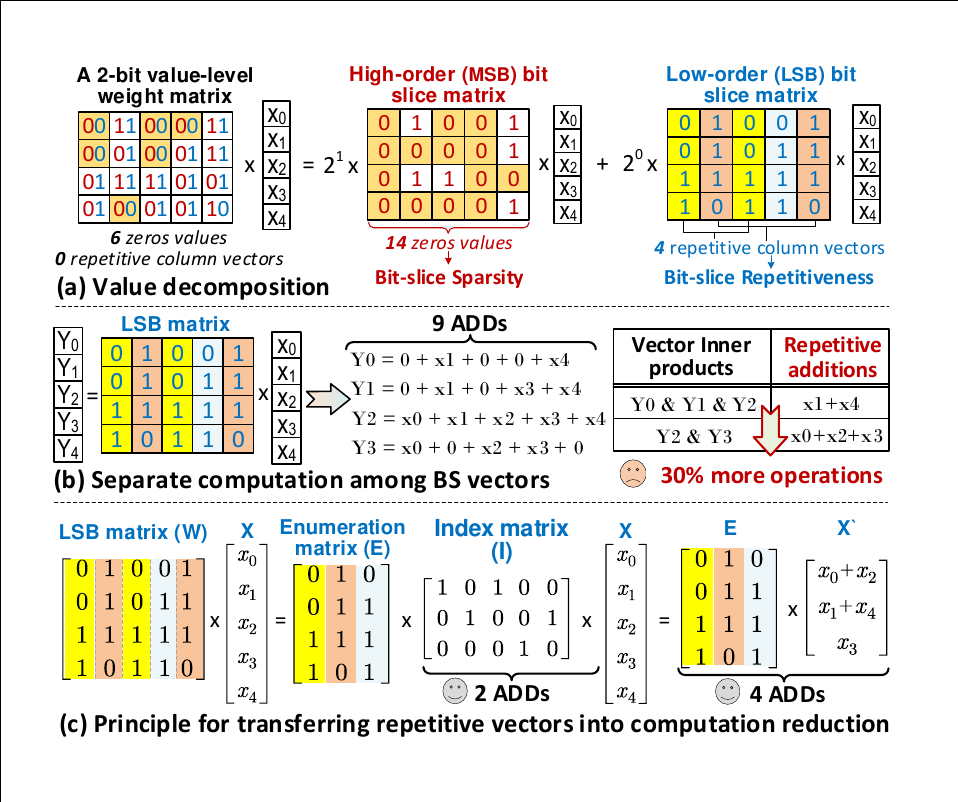}\vspace{-3mm}
\caption{Bit-level sparsity and repetition opportunities.}
\label{fig:Pigeonhole}\vspace{-2mm}
\end{figure}

Fig. \ref{fig:Pigeonhole} (a) shows that value-level representation obscures bit-level optimization opportunities. In the 2-bit value matrix, only six elements are zero, and no column vectors are repeated (Repeated column vectors can be used to accelerate GEMM). This is due to bit concatenation, where a $k$-bit zero requires all $k$ bits to be zero simultaneously. In contrast, decomposing the matrix into two 1-bit slices (MSB and LSB) reveals enhanced sparsity and redundancy. The MSB slice exhibits 14 zeros, yielding a $70\%$ sparsity rate (14/20). This aligns with the near-Gaussian distribution of weights \cite{liu2024spark,im2023sibia}, where higher-order bits tend to be zero. Additionally, the 1st and 2nd columns in the LSB slice are identical to the 3rd and 5th, respectively, indicating increased repetition after bit-level decomposition. Notably, a 2-bit integer GEMV is functionally equivalent to a shift-and-accumulate operation over the two bit-slice matrices, where the MSB slice is weighted by $2^1$ and the LSB by $2^0$. This demonstrates that bit-level decomposition preserves full compute equivalence while exposing fine-grained sparsity and redundancy. We refer to the above two opportunities as \textbf{BS sparsity} and \textbf{BS repetitiveness}.

However, directly computing with BS sparsity or repetitive vectors in a naive manner is still inefficient. With the LSB slice matrix in Fig. \ref{fig:Pigeonhole} (a) as an example, Fig. \ref{fig:Pigeonhole} (b) illustrates computing each BS vector independently. This results in redundant operations. Specifically, $x_1+x_4$ is calculated three times across $Y_0$, $Y_1$ and $Y_2$, while $x_0+x_2+x_3$ is recalculated twice, leading to a $30\%$ more operations. This inefficiency arises from failing to exploit redundancy across BS vectors. Naturally, this raises a key question: how can we harness such inherent repetitiveness to reduce overall computation?  

\begin{figure*}[t]
\centering
\includegraphics[width=\linewidth]{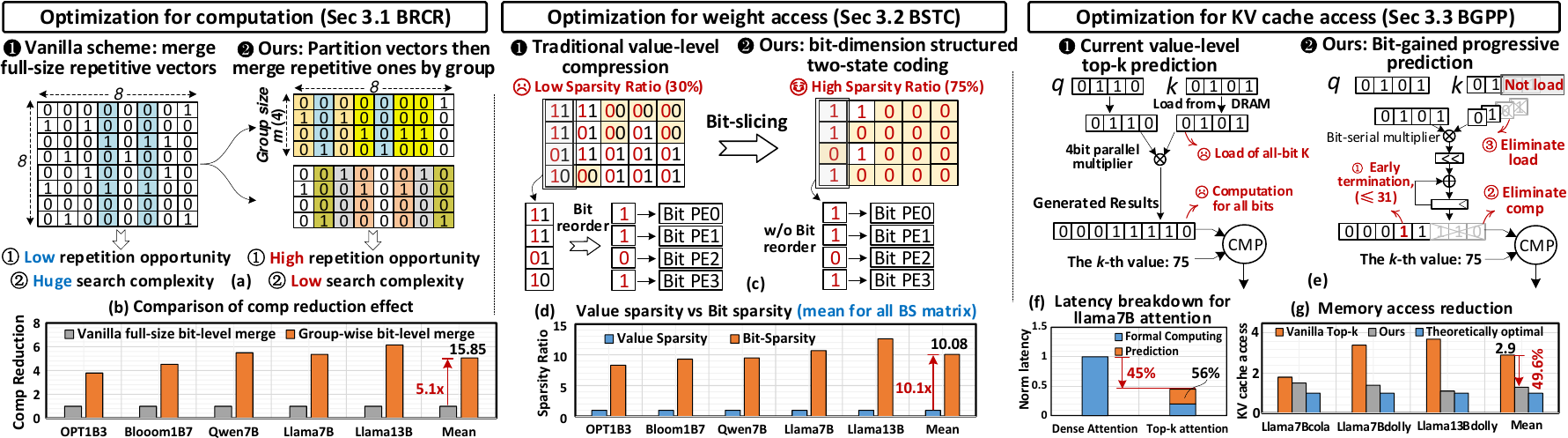}\vspace{-2mm}
\caption{Challenges and our strategies for applying bit-level computing to computation-memory-efficient LLM inference.}
\label{fig:Challeges}\vspace{-2mm}
\end{figure*}

\textbf{Opportunity}: Fig. \ref{fig:Pigeonhole} (c) illustrates an effective computation reduction strategy. First, it transforms the LSB matrix (denoted as $\mathbf{W}$) involving repetitive column vectors into an \emph{enumeration matrix} ($\mathbf{E}$) and an \emph{index matrix} ($\mathbf{I}$). Specifically, $\mathbf{E}$ stores unique column vectors from $\mathbf{W}$ and $\mathbf{I}$ records the mapping between each column in $\mathbf{W}$ and its corresponding vector in $\mathbf{E}$. For example, the 3rd column of $\mathbf{W}$ matches the 1st column of $\mathbf{E}$, so the value of $\mathbf{I}$ (1,3) is 1. This transformation rewrites  $\mathbf{W}\times\mathbf{X}$ as $\mathbf{E}\times\mathbf{I}\times\mathbf{X}$. The intermediate result $\mathbf{X'} = \mathbf{I} \times \mathbf{X}$ requires 2 additions, and $\mathbf{E} \times \mathbf{X'}$ requires 4 additions, yielding a $30\%$ reduction compared to the naive computation that demands 9 additions. We refer to this approach as a redundancy elimination strategy based on BS vector grouping. By grouping multiple BS vectors as a group matrix, the strategy identifies and eliminates redundant computations among them. Its effectiveness depends on the \textbf{repeated column vectors} within the group matrix: more repetition leads to lower computation.

\textbf{Challenges}: Despite its promise, designing an efficient bit-level accelerator for LLM inference remains a challenging task.


\vspace{2mm}

\hspace{-3.5mm}
\fbox{%
  \begin{minipage}{0.96\linewidth}
  \vspace{1pt}
\textbf{(Challenge 1)} Directly grouping a large number of BS vectors results in low repetition rates among column vectors.
  \vspace{1pt}
  \end{minipage}
}

\vspace{2mm}

As depicted in Fig. \ref{fig:Challeges} (a), the number of repetitive column vectors in an $8\times 8$ BS matrix is significantly smaller than the number of repetitive column vectors after decomposing it into two $4\times8$ sub-matrices, where we denote 4 as the group size $m$. This follows from the pigeonhole principle \cite{ajtai1994complexity}: When the number of holes is less than the number of pigeons, at least one hole will contain more than one pigeon. As  $m$ decreases, the number of available "holes" (i.e., at most $2^m$) is reduced, thus the probability of repetitive column vectors increases. Our analysis across five LLMs in Fig.~\ref{fig:Challeges}(b) reveals that, compared to the vanilla full-size merge, the group-wise merge achieves, on average, a $5.1\times$ reduction in computation.

\textbf{Key idea.} We decompose the large weight matrix $\in\mathbb{R}^{H\times H}$ in LLMs into several smaller group matrices of size $\mathbb{R}^{m\times H}$, where $H$ is hidden dimension. To support this, we design a CAM-based match unit, which significantly reduces the latency associated with the search process for repetitive column vectors along the H dimension.


\vspace{2mm}

\hspace{-3.5mm}
\fbox{%
  \begin{minipage}{0.96\linewidth}
  \vspace{1pt}
\textbf{(Challenge 2)} Value-level compression is incompatible with bit-level computation paradigms and obscures BS sparsity.
  \vspace{1pt}
  \end{minipage}
}
\vspace{2mm}

\begin{table}[t]
\renewcommand{\arraystretch}{1.1}
\caption{Summary for SOTA Transformer Accelerators.}\vspace{-4mm}
\begin{center}
\footnotesize
\begin{tabular}{l||m{0.94cm}<{\centering}|m{0.5cm}<{\centering}|m{0.5cm}<{\centering}|m{0.72cm}<{\centering}|m{0.63cm}<{\centering}|m{0.8cm}<{\centering}}
\specialrule{0.12em}{0.5pt}{0.4pt}
 \multirow{2}{*}{\!\!\!\textbf{Accelerator}} &   \multicolumn{2}{c|}{\textbf{GEMM}} & \multicolumn{2}{c|}{\textbf{Memory access}} & {\!\!\!\!\textbf{P $\&$ D}} & {\!\!\!\!\textbf{Optimiz.}}\\
\cline{2-5}
& \multirow{1}{*}{\!\!\!\!QKV $\&$ FFN}   & {\!\!\!Atten.} & {\!\!Weight}   &  \multirow{1}{*}{\!\!\!\!KV Cache} & \!\!\textbf{stage} & \!\!\textbf{Level} \\
\hline
\rowcolor{mygray}\!\!\!$\mathbf{A}^3$\! \cite{ham20203} & $\times$ & \checkmark  & $\times$ & $\times$ & P only & Value\\
\!\!\!\textbf{ELSA} \cite{ham2021elsa} & $\times$ & \checkmark  & $\times$ & $\times$ & P only & Value\\
\rowcolor{mygray}\!\!\!\textbf{Sanger} \cite{lu2021sanger} & $\times$ & \checkmark  & $\times$ & $\times$ & P only & Value\\
\!\!\!\textbf{DOTA} \cite{qu2022dota} & $\times$ & \checkmark  & $\times$ & $\times$ & P only & Value\\
\rowcolor{mygray}\!\!\!\textbf{DTATrans}\cite{yang2022dtatrans}\!\!\! &  $\times$ & \checkmark  & $\times$ & $\times$ & P only & Value\\
\!\!\!\textbf{Energon} \cite{zhou2022energon} & $\times$ & \checkmark  & $\times$ & Low & P only & Value\\
\rowcolor{mygray}\!\!\!\textbf{SpAtten} \cite{wang2021spatten} & \checkmark & \checkmark  & $\times$ & Low & P $\&$ D & Value\\
\!\!\!\textbf{SOFA} \cite{wang2024sofa} & $\times$ & \checkmark  & \checkmark & $\times$ & P only & Value\\
\rowcolor{mygray}\!\!\!\textbf{FACT} \cite{qin2023fact} & \checkmark & \checkmark  & Low & $\times$ & P only & Value\\
\!\!\!\textbf{MCBP} & {\fontsize{6.5}{5} \CheckmarkBold }& {\fontsize{6.5}{5} \CheckmarkBold }  & {\fontsize{6.5}{5} \CheckmarkBold } & {\fontsize{6.5}{5} \CheckmarkBold } & {\fontsize{6.5}{5} \textbf{P $\&$ D} } & \textbf{Bit} \\
\specialrule{0.12em}{0.5pt}{0.1pt}
\end{tabular}
\end{center}
\label{tab:works_comparision}\vspace{-5mm}
\end{table}

As shown in Fig. \ref{fig:Challeges} (c), this inefficiency stems from two primary factors: (1) Value-level compression achieves only a $30\%$ sparsity rate (SR), which is $2.5\times$ lower than bit-level sparsity; (2) Value-level formats require bit reordering for bit-level PEs, which incurs additional on-chip overhead. Fig. \ref{fig:Challeges} (d) further show that, across five LLMs, bit sparsity is on average $10.1\times$ higher than value sparsity.

\textbf{Illustration for the bit reorder}: Traditional memory layouts store multi-bit activations contiguously across bits. As a result, when computing the MSB slice, a large number of LSBs are also fetched unnecessarily. To enable bit-slice-based processing, a bit-reordering step is required to extract and reorganize the relevant MSB data into a contiguous MSB slice for input to the processing elements (PEs). We refer to this operation as bit-reorder.

\textbf{Illustration for the bit sparsity}: For a $k$-bit matrix, we first compute the bit sparsity of the bit-slice matrices for each bit position. The overall bit sparsity of the 8-bit matrix is then the average bit sparsity across all bit-slice matrices for each bit position.

\textbf{Key idea.} We propose an effective two-state coding scheme operating along the bit-slice dimension, naturally aligned with bit-level computation and eliminating the overhead of data reordering. The coding and computation are associated designed and operates at the same group granularity to ensure global maximum benefits. Lightweight en/decoders are designed to enable greater parallelism and low-power data coding within the same area budget.

\vspace{2mm}

\hspace{-3.5mm}
\fbox{%
  \begin{minipage}{0.96\linewidth}
  \vspace{1pt}
\textbf{(Challenge 3)} The current Top-$k$ prediction is coarse-grained and involves redundant computation and memory access.
  \vspace{1pt}
  \end{minipage}
}

\vspace{2mm}

As depicted in Fig.\ref{fig:Challeges} (f), although the top-$k$ prediction successfully reduces the overall attention latency by $45\%$, the bottleneck shifts to the prediction process itself. Therefore, it is imperative to further optimize the top-$k$ prediction process. Fig.\ref{fig:Challeges} (e) illustrates inefficiencies in value-based top-$k$ prediction using an example where the threshold is 75. To identify whether the current Key\,(0101) belongs to the top-$k$ set, the value-based approach loads the 4bit K entry from HBM and then executes computation for 8bit results. However, we observe that the top 2 bits alone are sufficient to determine that the final result ($\leq$31) will fall below the threshold, making the remaining 2-bit computation and memory access unnecessary.

\begin{figure}[t]
\centering
\includegraphics[width=\linewidth]{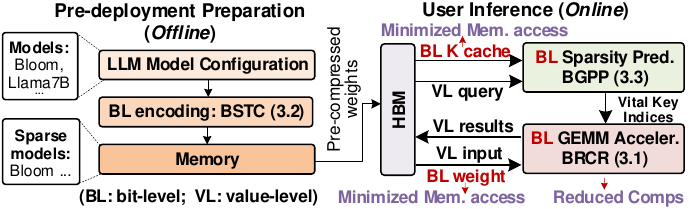}\vspace{-3mm}
\caption{The preparation and execution flow of MCBP.}
\label{fig:Deploy_flow}\vspace{-5mm}
\end{figure}

\begin{figure*}[t]
\centering
\includegraphics[width=\linewidth]{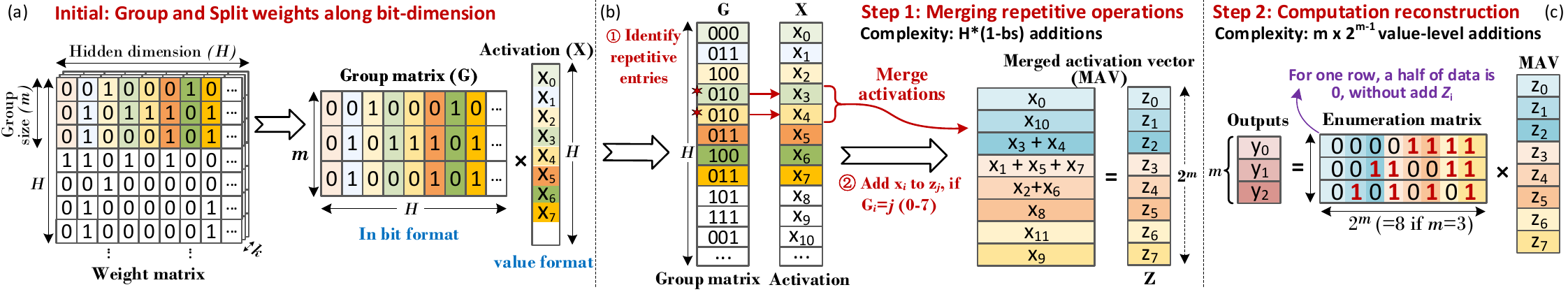}\vspace{-3mm}
\caption{Bit-slice-repetitiveness-enabled computation for GEMM (BRCR).}
\label{fig:bit_fuse_computing}
\end{figure*}

\textbf{Key idea}. We propose a bit-grained progressive prediction with early termination. Attention scores are estimated bit-wise from  MSB to LSB. This allows computation and KV cache access to be terminated early once the partial result exceeds the feasible top-$k$ range. As shown in Fig. \ref{fig:Challeges} (g), this reduces KV cache accesses by up to $50\%$ across three scenarios, compared to value-level prediction.

\textbf{Review of Transformer accelerators}. Unfortunately, current Transformer accelerators still struggle with computation and memory access issues,  due to their inability to exploit bit-grained opportunities for coordinated optimization. Table \ref{tab:works_comparision} summarizes their features. The majority of existing works \cite{ham20203,ham2021elsa,lu2021sanger,qu2022dota,yang2022dtatrans} focus on accelerating attention whose quadratic complexity dominates earlier encoder-based models, like BERT. However, their strategies are less effective for decoder-only LLMs during the autoregressive decoding stage, where performance is severely constrained by memory access. While Energon \cite{zhou2022energon} and SpAtten \cite{wang2021spatten} realize challenges with memory access, their coarse-grained pruning fails to handle fine-grained bit-level optimizations for both linear weights and KV cache. SOFA \cite{wang2024sofa} exhibits compute-memory co-optimization, but is restricted to attention. FACT \cite{qin2023fact} targets whole model computation reduction but lacks support for fine-grained KV cache and weight loading optimizations. These limitations motivate us to design an efficient LLM inference accelerator that jointly optimizes GEMM computation, weight access, and KV cache access across both prefill and decoding stages.

\section{Algorithm Optimizations of MCBP}\label{sec:BSO_mechanism}
Based on the three challenges, we propose three corresponding optimization strategies: BRCR, BSTC, and BGPP. Fig. \ref{fig:Deploy_flow} depicts the overall execution flow of MCBP. Model weights are offline-compressed into a bit-level (BL) sparsity format (BSTC, \S\ref{subsec:Weight access optimization}). During inference, the BL-compressed weights are loaded and decompressed, then sent for GEMM acceleration (BRCR, \S\ref{subsec:Bit_serial_Computation_Optimization}), the BL KV cache are on-demand fetched to predict attention sparsity (BGPP, \S\ref{subsec:KV Cache access optimization}). 

\subsection{BS-Repetitiveness-enabled Computation Reduction for GEMM (BRCR)}\label{subsec:Bit_serial_Computation_Optimization}
As depicted in Fig.\ref{fig:bit_fuse_computing} (a), the core idea of BRCR is first to decompose an $k$-bit weight matrix into $k$ bit-slice (BS) matrices. Then, for each BS matrix, it extracts $m$ rows of these matrices and merges them as a \emph{Group matrix}. Thus, it will process $m$ rows each time, instead of all rows. For clarity, we use GEMV to illustrate the acceleration mechanism, which is also effective in GEMM scenarios. Overall, two key steps are required to achieve computation acceleration.

\textbf{1) Merging repetitive operations}. 
As depicted by Fig. \ref{fig:bit_fuse_computing} (b), this step first \ding{172} identifies repeated entries (i.e., column vectors) in the \emph{Group matrix} $\mathbf{G}$, then \ding{173} merge their corresponding activations into a \emph{merged activation vector} (MAV), denoted as $\mathbf{Z}$. This is implemented by accumulating each activation into the partial sum of the corresponding entry in $\mathbf{Z}$, based on the value of each column in $\mathbf{G}$ (We denote as Grouped index). For example, the 3rd and 4th columns of the group matrix are both 010 (i.e. $G_{3}$=$G_{4}$=2), so their corresponding activations, $x_3$ and $x_4$ are added to the entry ($z_2$) of the $\mathbf{Z}$. Notably, for a bit column vector with $m$ elements, there are $2^m$ possible types. Thus, the MAV has a length of $2^m$. Mathematically, this step is equivalent to the $\mathbf{I}\times \mathbf{X}$ in Fig. \ref{fig:Pigeonhole} (c). Notably, non-zero entries in the MAV indicate multiple rows in a weight share the same addition operation. For instance, $z_3$ (Grouped index is $0\mathbf{11}$) denotes the repetitive additions among rows $1$ and $2$, while $z_0$ represents activations multiplied by zero, which can be directly eliminated. With bit sparsity ratio $bs$, this step consumes at most $H\times(1-bs)$ additions, regardless of group size $m$.   

\textbf{2) Computation reconstruction}. As depicted in Fig.\ref{fig:bit_fuse_computing} (c), this step is to reconstruct the GEMV results by multiplying the \emph{Enumeration matrix} with the MAV. It is noteworthy that for a Group matrix $\mathbb{R}^{m\times H}$, when \( H \) is very large, we can reasonably assume that all possible $2^m$ column vectors will appear. Thus, the Enumeration matrix contain all $2^m$ distinct column vectors. In this way, each row of the enumeration matrix can contain at most \( 2^{m-1} \) ones. Therefore, the computation reconstruction step requires at most \( m \times 2^{m-1} \) additions for reconstructing $m$-row GEMV.
 
In summary, for a $k$-bit, $m$-row GEMV with bit sparsity ratio $\tilde{bs}$ and value sparsity $vs$, where $\tilde{bs}$ is the average bit sparsity ratio across all ($\in[1,k]$) bit-slice matrices. The total additions required by BRCR is $k(H\times (1-\tilde{bs}) + m\times 2^{m-1})$. By contrast, existing sparsity-aware bit-serial computing (BSC) \cite{delmas2019bit,albericio2017bit} consumes $k(H\times m\times (1-\tilde{bs}))$ additions. And the value-based sparsity scheme consumes $H\times m\times k\times vs$ additions. For typical LLM models (H$\sim$4k, $\tilde{bs}$$\sim$0.70, vs$\sim$0.07, $m$=4), BRCR achieves up to $12.1\times$ and $3.8\times$ computation reduction compared to value sparsity and naive BSC. 

\textbf{Verify the existence for redundancy based on pigeonhole principle}. Any $m$-row binary matrix can have at most $2^m$ types of column vectors. Since LLMs (e.g., Bloom-7B, GPT-3) have hidden dimensions $H$ (4k–12k) far exceeding $2^m$, there are abundant opportunities for redundancy in LLMs.

\textbf{Key Insights: There is a key sweetspot of $m$ that achieves the maximum computation reduction while minimizing reconstruction overhead.} For a GEMV with a $k$-bit weight matrix $\in\mathbb{R}^{H\times H}$, the total operations of BRCR are \mbox{$kH^2/m\times(1-\tilde{bs})+kH2^{m-1}$}. The group size $m$ introduces an interesting trade-off. If $m$ is too small, it fails to exploit sufficient redundancy between the bit-slice vectors. Conversely, if $m$ is too large, the exponentially increasing reconstruction cost (i.e., $2^{m-1}$) offsets the benefits of redundancy removal. The DSE for optimal $m$ is provided in \S\ref{subsec:software_evalutation}.

\begin{figure}[t]
\centering
\includegraphics[width=\linewidth]{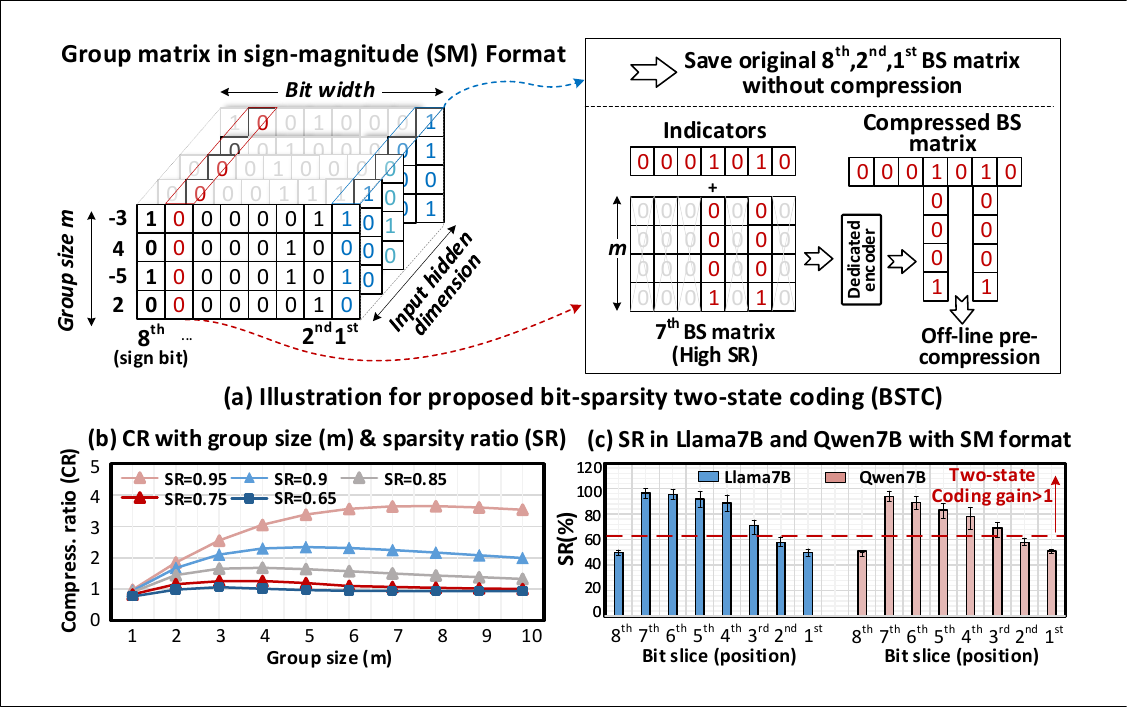}\vspace{-3mm}
\caption{BS-sparsity-enabled two-state coding (BSTC).}
\label{fig:huffman_coding}
\end{figure}

\subsection{BS-Sparsity-enabled two-state Coding (BSTC)}\label{subsec:Weight access optimization}

While numerous studies \cite{moons2016energy,han2015deep,moons201714,qin202523,han2016eie} have explored coding techniques for sparse weight compression, they largely focus on value-level sparsity, limiting their effectiveness. In contrast, BSTC exploits the key insight that quantized weights exhibit Gaussian-like distribution \cite{liu2024spark}, thus most non-zero weights own zero bits. To this end, BSTC encodes data at different BS matrices separately, to exploit the high sparsity in high-order bit plane. In addition, the encoding of BS matrices aligns with the computation granularity of BRCR, i.e., group size $m$, thus avoiding extra data conversion overhead.

Fig. \ref{fig:huffman_coding} (a) illustrates BSTC’s design. To exploit bit-0 sparsity in high-order bits (near MSB part), we adopt the sign-magnitude (SM) format for all weights. Given varying sparsity across bit positions, only bit-slice matrices from bits 3-7 are compressed, while bits 1, 2, and 8 remain uncompressed. Despite redundant sparsity in high-order bits, naively encoding would result in irregular data re-assignment for computation, leading to severe overhead. To this end, we employ a \emph{two-state} encoding, which distinguishes only zero data and non-zero data. Zero is encoded as 1'b0, and non-zero is encoded as a $(m+1)$-b symbol: $\{1'b1, m'b\,{\rm data}\}$. For instance, in Fig. \ref{fig:huffman_coding} (a), we have \{0000\}$\to$\{0\} and \{0001\}$\to$\{$\underline{1}0001$\}, where $\underline{1}$ is an indicator that facilitates decoding. In this way, BSTC provides regularity at the bit-column level and achieves lossless compression.  

\begin{figure}[t]
\centering
\includegraphics[width=\linewidth]{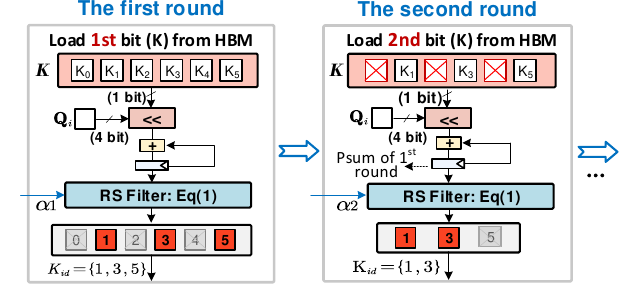}\vspace{-2mm}
\caption{Bit-grained progressive top-$k$ prediction (BGPP).}
\label{fig:filter}\vspace{-3mm}
\end{figure}

Since BSTC introduces a 1-bit indicator for each non-zero column vector, its applicability must be carefully evaluated; otherwise, the overhead may offset the encoding gains. Fig. \ref{fig:huffman_coding} (b) illustrates the compression ratio (CR) of BSTC under varying sparsity ratio (SR) as the group size ($m$) changes. There are some interesting insights: First, an excessively large $m$ may reduce the compression ratio due to fewer co-occurring zeros across data elements within larger groups. Second, when the SR is high, a larger group size $m$ tends to yield a higher compression ratio, as it reduces the relative overhead of storing indicators. Last, we can figure that when SR exceeds $65\%$, BSTC can achieve positive benefits (i.e. CR>1). Further, Fig. \ref{fig:huffman_coding} (c) analyzes the SR of bit-slice (BS) matrices across different bit positions in Llama7B and Qwen7B. It is observed that the SR for the 3rd to 7th BS matrices all exceed 65\%. Thus, we apply BSTC compression to these BS matrices. By contrast, for BS matrices with low SR, such as the 1st BS matrix, no compression is applied

\subsection{Bit-grained Progressive Prediction (BGPP)}\label{subsec:KV Cache access optimization}
As introduced in \S\ref{subsec:Sparsity_in_Atten}, the core idea of top-$k$ prediction is to estimate the attention matrix with a low-overhead paradigm, then pick up important Key indices. However, even utilizing the low-precision paradigm (e.g. 4bit with MSB only), the value-based strategy still causes unnecessary memory access and computation (Fig.\ref{fig:Challeges} (c)). Therefore, a more efficient prediction scheme is a must.

BGPP addresses this by leveraging the relative nature of softmax: if an input's gap from the current max exceeds a threshold, its softmax output will be near zero\cite{qin2023fact}. Thus, the gap (termed \emph{radius}) with the current max value can be used to filter trivial Keys. 

We propose a bit-grained progressive filter mechanism to achieve this. \emph{Progressive} means: it performs multiple rounds of filtering, where in each round, incremental filtering is applied based on the Keys (Ks) selected in the previous round. Fig. \ref{fig:filter} gives an illustration for this procedure. Assume the initial state consists of 6 Ks (${\rm K}_0$-${\rm K}_5$). In the first round, we fetch the MSB of all Ks for computation with $Q_i$ (with 4 bit), and obtain the estimated Max attention value denoted as max($\hat{A}_i^1$). Then, based on Eq.\eqref{eq:filter}, a radius-calculated (RS) filter obtains the filtering threshold for the current round. Then, it retains the indices ($K_{id}$) of the Ks (e.g. 1,3,5), whose attention values are greater than this threshold. In the next round, we only fetch the second bit of the $\{1,3,5\}$-th Ks from HBM. This process continues for the predetermined number of rounds.        

\begin{figure}[t]
\centering
\includegraphics[width=\linewidth]{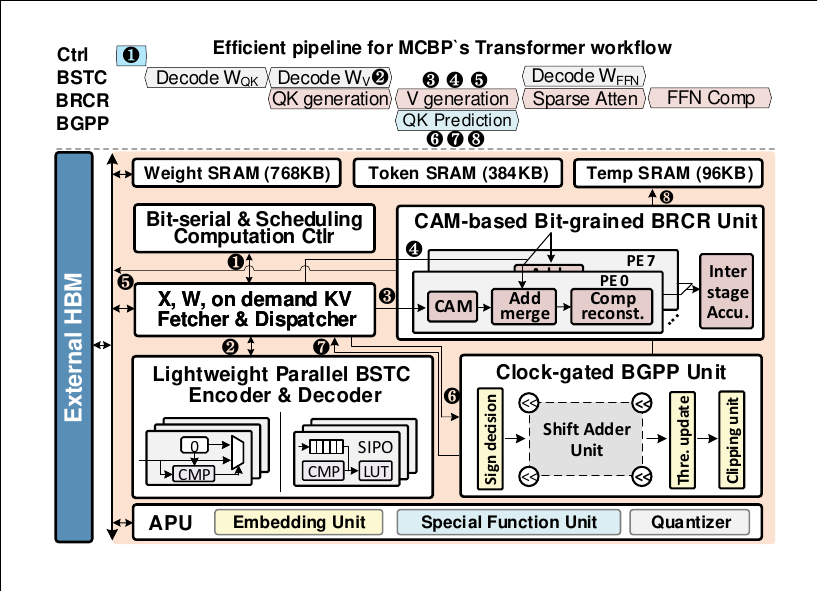}\vspace{-3mm}
\caption{High-level block diagram for MCBP accelerator.}
\label{fig:overall_architecture}
\end{figure}

Instead of directly adopting a fixed value as the threshold, for round $r$, we set the filter threshold of the $i$-th row as $\theta_i^r$: 
\begin{equation}
\theta_i^r=\max(\hat{A}_i^r)-\alpha_r\times radius, ~~~~0\le\alpha_r\le 1,
\label{eq:filter}
\end{equation}

\noindent where $\hat{A}_i^r$ is the estimated attention of the $i$-th row (During \emph{decoding} stage, $i$$=$$0$). Based on our experiments, we empirically set the default radius to $3$ and use a parameter $\alpha_r\in [0,1]$ to control the threshold. By adjusting $\alpha_r$, we can control the pruning ratio in each round.

\section{Architecture and Hardware Innovation}\label{sec:Overall_Architecture}

\subsection{Architecture Overview}\label{subsec:architecture_overview}

Fig. \ref{fig:overall_architecture} illustrates the MCBP architecture (bottom) and its workflow (top) for efficient Transformer inference under attention sparsity (\S\ref{subsec:Sparsity_in_Atten}). The accelerator operates through eight key steps, with numbered markers on the timeline indicating their positions within the overall pipeline. First, the controller sends token indices and bit-slice (BS) weights to the data fetcher, which decodes physical addresses and loads data into on-chip SRAM \ding{182}. The BSTC decoders then decompress the weight matrices \ding{183}, forwarding them to the BRCR unit \ding{184}, where a CAM-based module identifies repetitive BS weight entries.These indices are returned to the fetcher, translated into SRAM addresses, and used to fetch corresponding activations back to the BRCR unit \ding{185}. Finally, the computed GEMM results are written back to off-chip DRAM \ding{186}.

\begin{figure}[t]
\centering
\includegraphics[width=\linewidth]{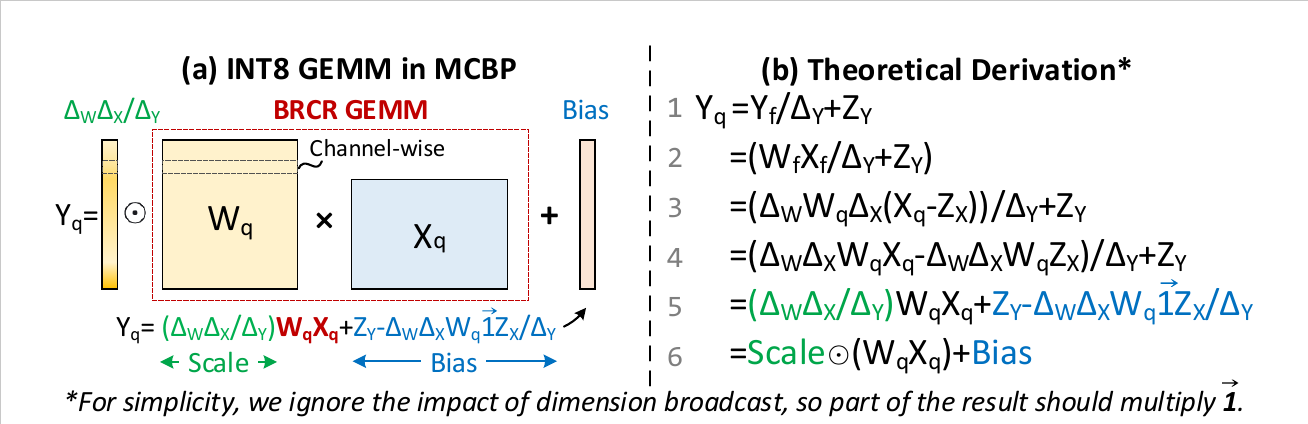}\vspace{-3mm}
\caption{Illustration for quantization process in MCBP.}
\label{fig:Zero_point}
\end{figure}

\begin{figure}[t]
\centering
\includegraphics[width=\linewidth]{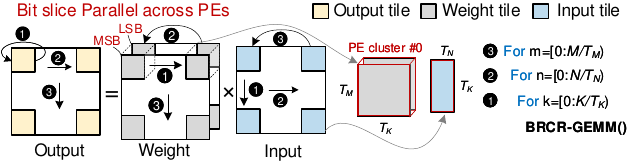}\vspace{-3mm}
\caption{The tiling strategy for GEMM in MCBP.}
\label{fig:MCBP_Tiling}
\end{figure}

To efficiently handle dynamic attention sparsity (\S\ref{subsec:Sparsity_in_Atten}) and hide prediction latency, BGPP operates concurrently with the BRCR unit. Its workflow begins by retrieving QK tensors from the \emph{data fetcher} \mbox{\ding{187}} and performs an initial prediction. The selected indices are then returned to the data fetcher \mbox{\ding{188}} to fetch the required Keys for the next round (as Fig.~\ref{fig:filter}). This process continues iteratively until the preset number of iterations is reached, and the final KV indices are stored in Temp SRAM \mbox{\ding{189}}. In this dedicated dataflow, once computation proceeds to the attention part, the BRCR unit can merely calculate attention scores with those vital KVs, based on these KV indices generated by BGPP. To fully support Transformer computation, MCBP integrates an Auxiliary Processing Unit (APU) that includes an embedding unit for generating input token embeddings via table lookup, a special function unit (SFU) implemented in FP16 using a combination of lookup tables and polynomial approximation \mbox{\cite{qin202523}} to compute non-linear functions such as GELU, softmax, and layer normalization, and a quantizer that handles data conversion between FP16 and INT8. Notably, the concatenation in MHA is performed during data movement.

\textbf{Processing of scaling and zero point}. 
Fig. \ref{fig:Zero_point} (a) illustrates the quantization process in MCBP, where weights are quantized using per-channel symmetric quantization, and activations are quantized using per-tensor asymmetric quantization, as \mbox{\cite{kam2024panacea,xiao2023smoothquant,wei2022outlier}}. Taking activations as an example, the quantized input activation is computed as $X_q$=$X_f/\Delta_x$$+$$Z_x$, where $\Delta_x$ is the scaling factor and $Z_x$ is the zero-point offset. Notably, $\Delta_w$, $\Delta_x$, $\Delta_y$, $Z_x$ and $Z_y$ can be pre-known by the calibration dataset. Based on the derivation in Fig. \ref{fig:Zero_point} (b), the final output is expressed as $Y_q=Scale \odot (W_qX_q)+Bias$, where the INT GEMM $(W_qX_q)$ is accelerated by the BRCR unit via efficient bit-slice processing and shift-accumulation. The results are then processed by the quantizer with the scaling and bias terms.

\textbf{Tiling Strategy.} Fig. \ref{fig:MCBP_Tiling} illustrates the tiling strategy of MCBP and its corresponding loop representation for the output-stationary dataflow. To maximize weight reuse, MCBP stores slices included in the $T_M$ $\times$ $K$ weight tile into the weight SRAM at once, if possible. Then the BRCR unit assigns a $T_M \times T_K$ weight tile together with a $T_K \times T_N$ activation tile to each PE cluster, where 8 PEs concurrently process each bit-slice of the weight tile in parallel. In this work, we set $T_M = 64$, $T_K = 256$, and $T_N = 32$. 

\begin{figure}[t]
\centering
\includegraphics[width=\linewidth]{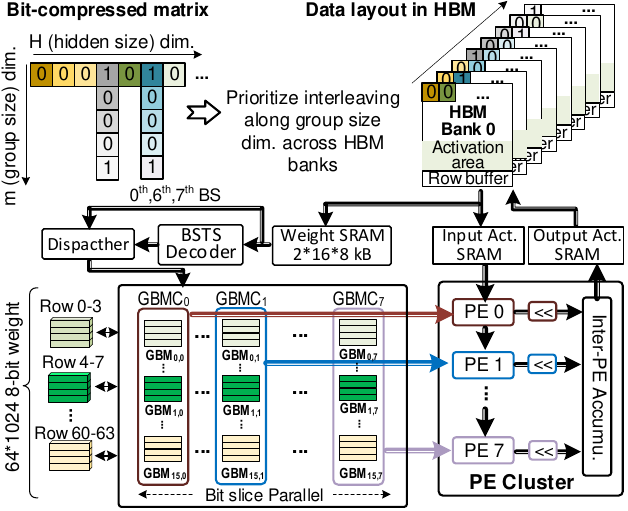}\vspace{-2mm}
\caption{The bit-grained computation dataflow of MCBP.}
\label{fig:Dataflow}
\end{figure}

\subsection{Bit Dataflow of MCBP}\label{subsec:dataflow}
To optimize bit-level memory access, we orchestrate the BS weight matrix layout in off-chip HBM. Fig. \ref{fig:Dataflow} shows an example using the compressed weight from Fig. \ref{fig:huffman_coding}. Given HBM’s read-write characteristics, we prioritize storing the bits along the group size dimension at the same address across all banks. Once filled, allocation moves sequentially to the next address until the entire BS matrix is stored. Before computation, the BS data is loaded into the on-chip weight SRAM.  Given the one-row-per-cycle access feature of SRAM banks, we prioritize storing the BS matrix within a single bank. Once the data in the weight SRAM is ready, the sparse BS matrices are sent to the BSTC decoder for decompression. During the process, the controller will check whether decoding is required. BS matrices at the 1st, 2nd, and 8th bit positions are not decoded, as they were not encoded due to their low sparsity (see Fig. \ref{fig:huffman_coding}).

\subsection{CAM-based BRCR Unit}\label{subsec:CAM_Acceleration}
The BRCR mechanism requires quickly identifying and consolidating identical elements in the group matrix (\S \ref{subsec:Bit_serial_Computation_Optimization}). To this end, we design a content-addressable memory (CAM) based fast match unit, which can identify identical elements in one cycle.

\begin{figure}[t]
\centering
\includegraphics[width=\linewidth]{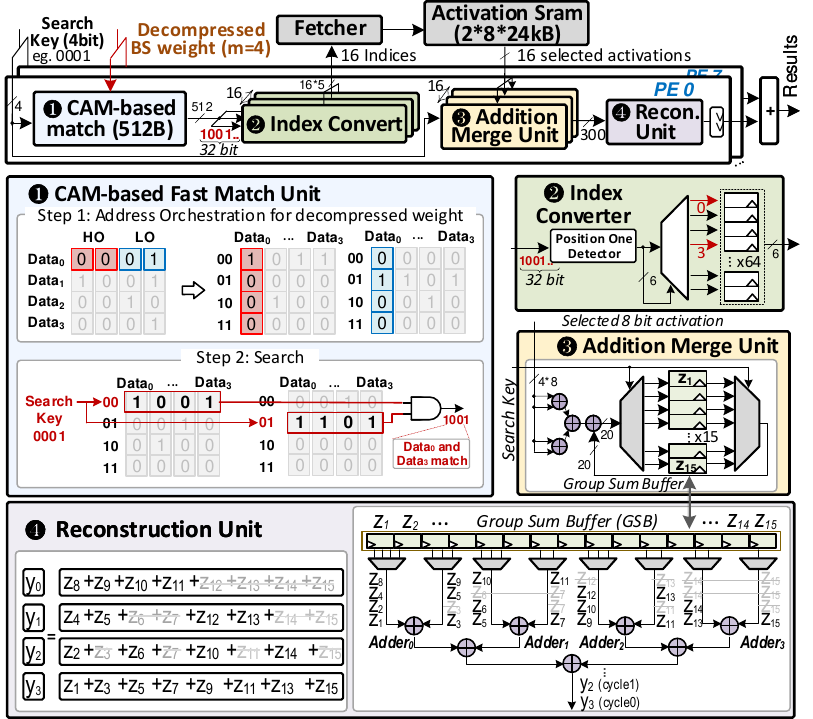}\vspace{-2mm}
\caption{A PE cluster of CAM-based fast-match BRCR unit.}
\label{fig:Adder_group}
\end{figure}

As depicted in Fig. \ref{fig:Adder_group} \ding{182}, we adopt a group size $m=4$. 
Initially, each 4-bit column vector of the decoded weights is orchestrated in CAM. Higher-order (HO) two bits and lower-order (LO) two bits of each 4-bit data are managed separately. As two bits correspond to four possible values, four entries are needed to store the orchestrated data. Then, for the search step, if an entry address matches the search key, the content of that entry is set to 1, while the other entries are set to 0. Taking searching 0001 (search key) as an example, the HO two bits read the row at address `00' of the MSB bank, while the LO two bits read the row at address `01' of the LSB bank. Then readout bits from both banks are ANDed to match both high and low 2 bits with the search key. The generated bitmap ‘1001’, indicates $x_0$ and $x_3$ match the 0001. The controller enumerates all possible search keys for $m$=4 (0000 to 1111). If the search key is 4'b0000, the CAM will be clock-gated to save power. The CAM, with a 2-bit matching length as its basic block, is designed to be reconfigured by re-matching the outputs of multiple basic blocks, to support adaptation to different group sizes.

The CAM-generated bitmap identifies activations to be merged and added. Sixteen index converters then translate the bitmap (e.g., 1001) into corresponding activation indices (\ding{183}). Next, the search key (0001) and the fetched activations ($x_0$, $x_3$) are together sent to the addition merge unit (AMUs) (\ding{184}). The AMU first adds $x_0+x_3$, then put the psum to the first register $z_1$, based on the search key 0001. For more fetched activations, data in one register is read out by the MUX and added to the psum, then the result is written back to the same register in group sum buffer (GSB) by the deMUX.

Next, a reconstruction unit (RU) reorganizes partial sums stored in GSB into correct results. Inspired by the fixed re-construct formula as Fig. \ref{fig:Adder_group} (\ding{185}) left, we design a low-power RU with a fixed data path. \emph{Fixed} means: we bind specific registers to inputs of each adder. By reordering the computation sequence, we extend the data lifecycle in adders. For example, computing $y_3$ first, followed by $y_2$ down to $y_0$, allows \emph{Adder 3} to read $z_{15}$ only once, reducing its switching activity by up to $75\%$. Given that the reconstruction workload is much lighter than the addition merging, one RU is time-multiplexed to serve 16 AMUs, improving resource utilization.

\subsection{Lightweight BSTC CODEC and Data Layout}
\begin{figure}[t]
\centering
\includegraphics[width=\linewidth]{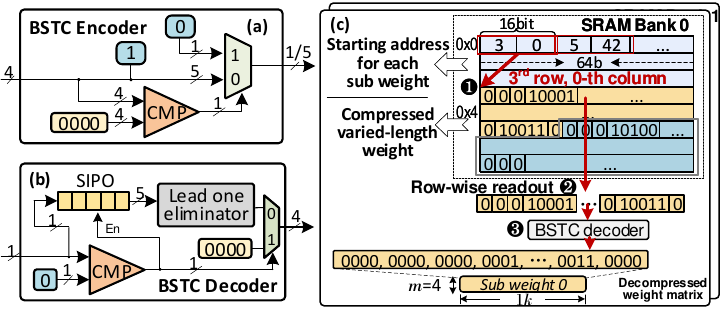}\vspace{-2mm}
\caption{(a)(b) Architectures for lightweight BSTC encoder/decoder. (c) Parallel-friendly segmented data layout.}
\label{fig:Huffman_hardware}
\end{figure}

We first design a lightweight encoder-decoder that enhances parallelism within the same area budget. Then, we introduce a segmented interleaved data layout in SRAM to support parallel en/decoding. Fig.\ref{fig:Huffman_hardware} (a)(b) depicts the lightweight BSTC en/decoder architectures. The encoder comprises a 4 bit comparator (CMP) and a MUX. If the input is non-zero, it adds a 1-bit `1' ahead the MSB and outputs the result; Otherwise, it outputs a 1-bit `0'. The decoder includes a 1-bit CMP, a 5-bit serial-in parallel-out (SIPO) register (for $m$=4), and a leading one eliminator. When a `0' is detected in the bit stream, it outputs four consecutive 0's. Otherwise, it buffers received bits in the SIPO, which outputs the buffered content once full.

Fig.\ref{fig:Huffman_hardware} (c) illustrates the segmented weight layout during a decompression process. To enable parallel decoding, the weight matrix is partitioned along the hidden dimension into multiple sub-weights, each stored independently in separate banks. Given variable compression ratios, the starting address of each sub-weight is recorded. Before decompression, the controller fetches these starting addresses from the address area \ding{182}. Based on the retrieved addresses \ding{183}, sub-weight data is accessed row-wise and sent to the BSTC decoder \ding{184} for decompression. Each bank has 64 columns and 1024 rows, we use a 6-bit column address and a 10-bit row address to locate the first data of each sub-matrix. One row can store the address of four 1k-length sub-matrices, and three rows suffice to index up to 12 sub-matrices, covering the weight size of most LLMs. \vspace{-1mm}

\subsection{Threshold-aware clock-gated BGPP Unit}\label{subsec:KV_Cache Filter}
\begin{figure}[t]
\includegraphics[width=0.99\linewidth]{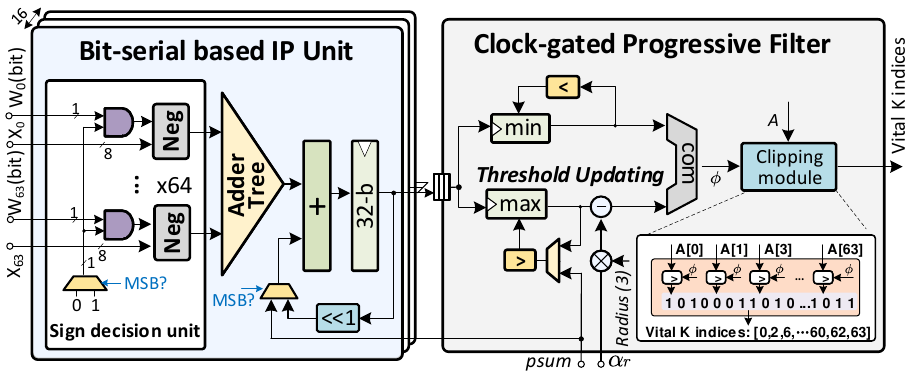}\vspace{-3mm}
\caption{Threshold-aware clock-gated BGPP unit.}
\label{fig:Progressive_Filter}\vspace{-3mm}
\end{figure}

Fig. \ref{fig:Progressive_Filter} shows the architecture of BGPP unit. First, 16 bit-serial inner product units compute $\mathbf{Q}$ (1*64) $\times$ $\mathbf{K}$ (64*16), each with a 64-input adder tree. The generated summations are passed to the progressive filter (PF) unit. Next, a \emph{threshold updating} (TU) module serially identifies the Max and Min values from these data. After completing the statistics for one row of the estimated attention, the TU subtracts the $\alpha_r\times radius$ from the Max value, which stands for the filter threshold (Eq.\eqref{eq:filter}). Next, a \emph{clipping} module compares all attention entries with this threshold, then produces a binary mask signal, where ``1" signifies the index of Ks eligible to proceed to the subsequent filtering round. After a fixed number of rounds, the final set of key indices is selected. To save power, if the threshold falls below the observed minimum, the clipping module is clock-gated and BGPP immediately proceeds to the next round. Additionally, to enable bit-wise computation under the SM format, we design a \emph{sign decision unit (SDU)} and place it before the adder tree. \vspace{-2mm}

\section{Evaluation}\label{sec:evaluation}
\subsection{Experimental Setup}\label{subsec:Experimental_setup}
\textbf{Baseline comparisons}: We compare MCBP with two SOTA bit accelerators: FuseKNA \cite{yang2021fusekna}, Bitwave \cite{shi2024bitwave}, and three Transformer accelerators: Spatten \cite{wang2021spatten}, SOFA \cite{wang2024sofa}, FACT \cite{qin2023fact}. For fair comparison, FuseKNA and Bitwave are adapted from convolution to GEMV using im2col. All designs are normalized to a 28nm process and evaluated under identical conditions: PE arrays occupy the same area as MCBP and work in 1GHz, on-chip SRAM is set to 1248kB, and HBM bandwidth is fixed at 512-bit/cycle, with 4 pj/bit \cite{o2017fine}.

\begin{table*}[h]
\scriptsize
\renewcommand{\arraystretch}{1.05}
\caption{Accuracy of Different Language Models with FP16, INT8 and MCBP Optimization (S: Standard, A: Aggressive).}\vspace{-4mm}
\begin{center}
\begin{tabular}{l|m{0.3cm}<{\centering}m{0.45cm}<{\centering}m{0.35cm}<{\centering}m{0.35cm}<{\centering}m{0.5cm}<{\centering}m{0.3cm}<{\centering}m{0.3cm}<{\centering}m{0.25cm}<{\centering}m{0.45cm}<{\centering}m{0.45cm}<{\centering}m{0.4cm}<{\centering}m{0.4cm}<{\centering}m{0.55cm}<{\centering}m{0.3cm}<{\centering}m{0.3cm}<{\centering}m{0.3cm}<{\centering}m{0.5cm}<{\centering}m{0.4cm}<{\centering}m{0.5cm}<{\centering}m{0.4cm}<{\centering}m{0.5cm}<{\centering}c}
\specialrule{0.12em}{0.5pt}{0.8pt}
\!\!\textbf{Model} & \multicolumn{8}{c}{LlaMa7B} & \multicolumn{8}{c}{LlaMa13B} & \multicolumn{2}{c}{OPT1B3} & \multicolumn{2}{c}{Bloom1B7} & \multicolumn{2}{c}{Qwen7B} \\
\hline 
\!\!\textbf{Task}$^{\ddagger}$ & \!\!\!MMLU & \!\!\!Wikiling. & MBPP & \!Wiki2 & \!\!\!\!Winogran. & Cola & \!\!MNLI & \!\!SST2 & \!\!MMLU & \!\!Wikiling. & MBPP & Wiki2 & \!\!Winogran. & Cola & \!\!MNLI & SST2 & \!\!Wikiling. & MBPP & \!\!Wikiling. & MBPP & \!\!Wikiling. & MBPP \\
\!\!\textbf{FP16} & 35.1\% &  39.3 & $17.8\%$ & 5.68 & 70.1\% & 80.3\% & 84.6\% & \!\!92.5\% & 41.2\% & 43.3 & 22\% & 5.09 & 74.2\% & 82.5\% & 85.5\% & \!93.8\% &  36.2 & 12\% & 44.3 & 16\% & 46.6 & 30\%   \\
\!\!\textbf{INT8} & 34.7\% & 38.9 & 17.2\% & 5.73 & 69.3\% & 80.2\% & 84.4\% & \!\!92.5\% & 40.9\% & 42.7 & 21.6\% & 5.13  & 73.7\% & 82.3\% & 85.3\% & \!93.7\% & 35.9 & 11.6\% &  44.1 & 15.7\% & 46.4 & 29.2\%   \\
\!\!\textbf{MCBP (S)}\!\! & 34.6\% & 38.8  & 17.1\% & 5.75 & 69.2\% & 80.2\% & 84.4\% & \!\!92.4\% & 40.7\% & 42.6 & 21.5\% & 5.15 & 73.4\% & 82.3\% & 85.3\% & \!93.7\% & 35.8 & 11.5\% & 44.0 & 15.6\% & 46.3 & 29.1\%\\
\!\!\textbf{MCBP (A)}\!\! & 34.1\% & 38.4 & 16.5\% & 5.80 & 68.7\% & 80.0\% & 84.1\% & \!\!92.1\% & 40.2\% & 42.0 & 21.0\% & 5.19 & 72.8\% & 82.0\% & 85.1\% & \!93.4\%  & 35.3 & 11.0\% &43.6 & 15.2\% & 45.9 & 28.4\%\\
\specialrule{0.12em}{0.5pt}{0.5pt}
\end{tabular}\vspace{-1mm}
\begin{flushleft}
$^{\ddagger}$ MMLU, WinoGrande, MBPP, Cola, MNLI, SST2 are evaluated by accuracy. Wikitext2 is evaluated by perplexity, where lower is better. Wikilingua is evaluated by ROUGE-1, where higher is better. 
\end{flushleft}
\end{center}
\label{tab:accuracy}
\end{table*}

\textbf{Benchmarks}: We evaluate MCBP on several LLM models of varying sizes, including Llama7B/13B \cite{touvron2023llama2}, Qwen7B \cite{bai2023qwen}, Bloom1B7 \cite{le2022bloom} and OPT1B3 \cite{zhang2022opt}, across nine tasks. These tasks includes Cola (S=0.25k), MNLI (S=0.5k), SST2 (S=0.25k) from GLUE \mbox{\cite{wang2018glue}}, language modeling (Wikitext-2 (S=2k) \mbox{\cite{merity2016pointer}}, Wikilingua (S=2k) \mbox{\cite{ladhak-wiki-2020}}, Winogrande (S=0.25k) \mbox{\cite{sakaguchi2021winogrande}}), Multitask Language Understanding (MMLU, S=0.5k)  \mbox{\cite{hendrycks2020measuring}}, code generation MBPP (S=1k) \mbox{\cite{austin2021program}}, long
context processing dolly (S=8k) \mbox{\cite{conover2023free}}.

\textbf{Quantization Accuracy}. All pre-trained models are sourced from Pytorch \cite{paszke2017automatic} and HuggingFace \cite{wolf2020transformers}. INT8 baselines derived via post-training quantization, where only the GEMMs are quantized to INT8, while non-linear operators (e.g., softmax) remain in FP16 precision. As shown in Table. \ref{tab:accuracy}, \textbf{the INT8 baseline incurs less than a $1\%$ average accuracy drop from FP16, confirming its validity}. Notably, for reasoning tasks such as MMLU and Winogrande, the accuracy degradation caused by INT8 quantization is negligible, typically below $0.5\%$. This observation is consistent with prior works \mbox{\cite{jacob2018quantization}}, which suggests that classification and reasoning tasks, due to their discrete output space and robustness to quantization noise, exhibit a high tolerance for low precision.

\begin{figure*}[t]
\centering
\includegraphics[width=\linewidth]{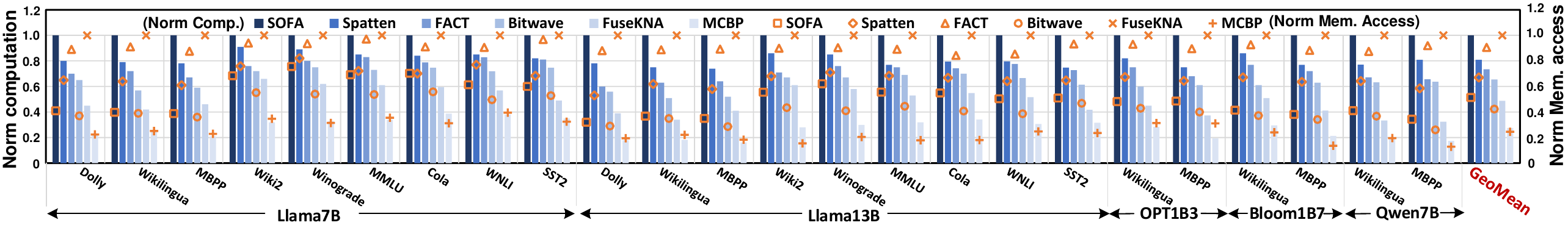}\vspace{-4mm}
\caption{Normalized computation (\emph{prefill} stage) and memory access (\emph{decoding} stage) of LLM inference.}
\label{fig:Overall_computation}
\end{figure*}

\begin{figure}[t]
\centering
\includegraphics[width=0.97\linewidth]{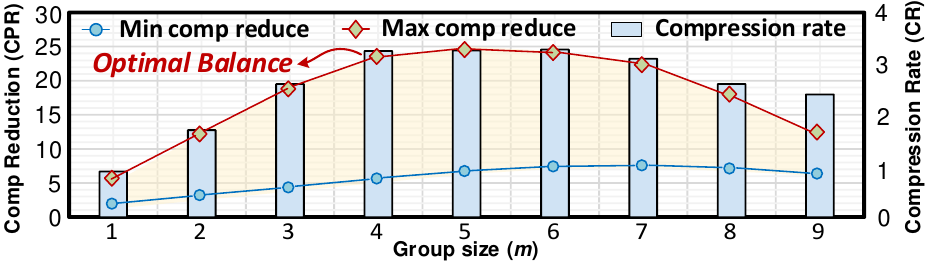}\vspace{-2mm}
\caption{Design space exploration of the optimal group size $m$, for computation reduction and compression rate.}
\label{fig:m_selection}
\end{figure}

\textbf{Simulation}: We implement the RTL design for MCBP and utilize Synopsys DC on TSMC 28nm CMOS technology to estimate the logic area and power. The CAM cell is designed using Cadence Virtuoso at the schematic level, then integrated with Verilog-based digital peripherals. The power, area, and read/write bandwidth of on-chip SRAM buffers are estimated through CACTI \mbox{\cite{muralimanohar2009cacti}}. Off-chip HBM modeling involves simulating row activation and access patterns under various data layouts, capturing HBM’s burst behavior. We derive memory latency from Ramulator \mbox{\cite{kim2015ramulator}}, and estimate IO power following the methodology in~\mbox{\cite{cavigelli2016origami,andri2016yodann,wang2017energy}}. We extract each stage’s cycles by simulating the RTL with Verilator~\mbox{\cite{snyder2004verilator}}, and use a custom cycle-level simulator to evaluate end-to-end performance.

\textbf{GPU comparison} We run benchmarks on Nvidia A100 with SOTA TensorRT-LLM \mbox{\cite{Nvidia2023}}. To exclude the software overhead, we measure execution time with \emph{cudaEvent}, isolating GPU execution from CPU interference. The GPU is dedicated during tests, and large batch sizes are used to amortize data transfer costs. We employ \emph{nvprof} to exclude non-computational phases. Power is measured via \emph{nvidia-smi}; dynamic power is computed as the difference between active and idle states. Each experiment is run 2k times, discarding the top and bottom $15\%$ before averaging.

\subsection{Algorithm Performance}\label{subsec:software_evalutation}
\textbf{Algorithm settings}: We regard INT8 models as the accuracy baseline, and adjust the value of $\alpha_r$ in 0.1 increments to evaluate the accuracy and overhead for each benchmark. This yields two MCBP configurations: standard ($0\%$ loss), aggressive ($1\%$ loss), representing the minimal and maximal performance optimizations, respectively.

\textbf{Optimal Group Size}. We determine the optimal group size $m$ by comparing computation reduction (CPR) and compression rate (CR) against dense models. Considering the varying sparsity levels, both the max and min CPRs are reported. Fig. \ref{fig:m_selection} shows that CPR raises from $m$=1 to $m$=5, as more weight rows are merged, but declines beyond $m$=5, due to the exponential growth ($2^m$, Fig. \ref{fig:bit_fuse_computing}) in additions required by computation reconstruction. For CR, $m$=1 results in a CR of less than 1, while $m$=4 maximizes CR by capturing all-zero columns. Beyond this point, fewer all-zero columns, in turn, negatively impact the CR. Considering the balance CPR and CR, and that 4 is the common divisor of most Transformer hidden dimensions, we select $m$=4 for this work.

\textbf{Computation Reduction}. Fig. \ref{fig:Overall_computation} compares the computation reduction of LLM inference across different accelerators. SOFA, which focuses solely on attention and adopts coarse-grained value-level sparsity, yields the lowest reduction and is used as the baseline. Bitwave enhances performance by exploiting bit-level sparsity, achieving a $32\%$ reduction, and outperforming value-sparsity-based accelerators like FACT and Spatten. However, it does not capitalize on bit-repetition. FuseKNA utilizes bit-repetition but fails to exploit attention sparsity, limiting its reduction to $49\%$. By contrast, MCBP achieves up to $72.4\%$ reduction by exploiting fine-grained bit-repetition, sparsity and attention dynamic sparsity. 

\begin{figure}[t]
\centering
\includegraphics[width=\linewidth]{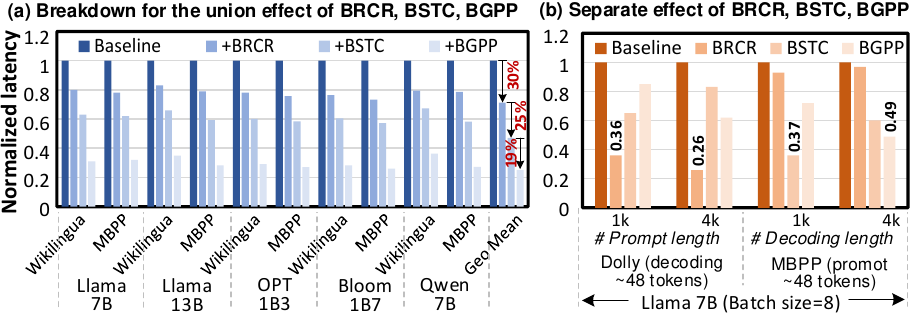}\vspace{-2mm}
\caption{Latency reduction for BRCR, BSTC and BGPP.}
\label{fig:latency_break_down}\vspace{-6mm}
\end{figure}

\textbf{Memory Access Reduction}. FuseKNA, which only exploits value compression by run-length coding, serves as the baseline. FACT and Spatten utilize IO-intensive top-$k$ to speculate trivial KV tokens, which successfully reduces computation but leading to redundant IO traffic. SOFA exclusively targets KV memory traffic in the attention module via cross-stage tiling, but does not mitigate weight traffic during the \emph{decoding} stage. Thus, it shows comparable memory reduction to Bitwave in long-sequence tasks (e.g., Dolly, Wikilingua), but performs less effectively with short sequences like Cola. This is because, in short-sequence tasks, the memory bottleneck lies in the weight traffic, which SOFA fails to mitigate. In contrast, MCBP achieves an average memory reduction of $75.8\%$ across both long- and short-sequence tasks, attributed to BSTC and BGPP, which respectively reduce weight and KV cache traffic.

\subsection{Architecture Evaluation}\label{subsec:Architecture_evalutation}

We first set an ablation study to evaluate the latency reduction of BRCR, BSTC and BGPP against a baseline, which is assumed to be vanilla bit computation + value-level Huffman compression + value-level top-$k$ prediction. Latency is evaluated by mapping various workloads on the MCBP accelerator. As shown in Fig.\ref{fig:latency_break_down} (a), BRCR reduces average latency by $30\%$ over the baseline, by eliminating redundant computation in the prefill stage. Further, BSTC and BGPP achieve a further $44\%$ latency reduction by significantly reducing I/O traffic from weights and KV cache during decoding.

\begin{figure}[t]
\centering
\includegraphics[width=\linewidth]{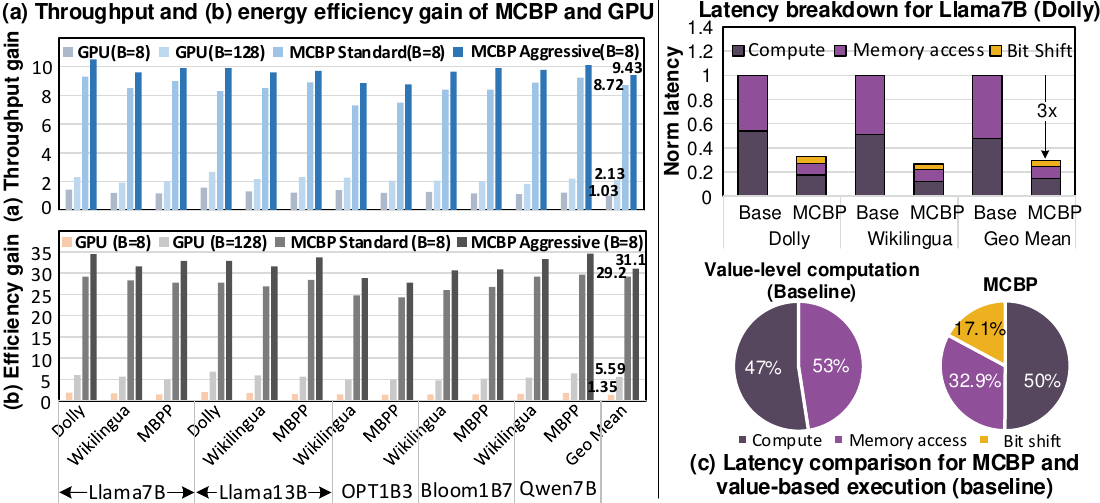}\vspace{-3mm}
\caption{(a) Throughput and (b) energy efficiency gain of MCBP. (c) Breakdown for the overhead of bit shifting.}
\label{fig:GPU_gain}
\end{figure}

Fig.\ref{fig:GPU_gain} (c) profiles the latency overhead between typical value level INT8 computation and MCBP (bit-level). Despite a $17\%$ bit-shifting overhead, the overall $3\times$ latency reduction proves that the gain achieved through bit sparsity effectively covers this overhead.

Fig.\ref{fig:latency_break_down} (b) shows the individual contributions of BRCR, BSTC, and BGPP across two LLaMA-7B tasks. For the Dolly long-text summarization task, we maintain a decoding length of 48 tokens and test different schemes with varying prompt lengths. In this case, BRCR delivers the primary speedup, achieving $3.9\times$ and $2.8\times$ latency reduction for 1k and 4k prompts, respectively, while BSTC and BGPP achieve only $1.6\times$ and $1.2\times$ acceleration at 1k prompt. This is because GEMM computation dominates $55\%$ of total latency in prompt-driven long-text summarization, making BRCR the most effective. With 4k prompts, BGPP outperforms BSTC due to increased KV cache memory access. For code generation task MBPP, BRCR only reduces latency by $1.2\times$, as the serial autoregressive decoding stage dominates latency. With 1k decoding length, BSTC achieves $2.7\times$ latency reduction for weight traffic reduction, and BGPP achieves $1.4\times$ for KV cache reduction. With $4k$ decoding length, BGPP increases to $2.1\times$, while BSTC drops to $1.6\times$. 

\emph{Throughput Gain:} Fig. \ref{fig:GPU_gain} (a) compares the throughput of MCBP with A100 GPU on all benchmarks with batch sizes of 8 and 128. Given the INT8 compute power of A100 is 624 TOPS, we use 148 MCBP processors (total with 622TOPS@INT8) with data and model parallelism for performance comparison. First, we observe that B=128 provides an average $2.1\times$ throughput gain over B=8, primarily due to amortized memory access. However, this benefit saturates, as a $16\times$ increase in batch size results in only a $2\times$ throughput gain. In contrast, MCBP achieves an $8.72\times$ speedup over A100 with the same batch size. Second, we can see naively applying MCBP's algorithm on GPU yields only 1.03$\times$ speed up, as GPUs cannot exploit bit-slice repetition and fine-grained dataflow and progressive sparsity prediction. In contrast, MCBP accelerator achieves 78$\%$ average utilization due to its Transformer-oriented workflow, which fully pipelines Parallel BSTC en/decoders, BRCR acceleration, BGPP predictor, leading to nearly $8\times$ higher sparsity utilization than GPU. Overall, MCBP standard and aggressive achieve an average $8.72\times$/$9.43\times$ inference speed up, respectively.

\begin{figure}[t]
\centering
\includegraphics[width=\linewidth]{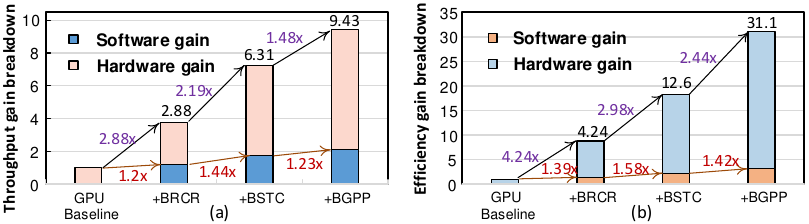}\vspace{-4mm}
\caption{Throughput and energy efficiency gain breakdown.}
\label{fig:Gain_breakdown}
\end{figure}

Fig. \ref{fig:Gain_breakdown} (a) gives the throughput gain breakdown, where \textbf{software gain} refers to the improvements achieved by directly deploying software optimizations on the GPU. Although the bit repetition-leveraged BRCR theoretically reduces computation by $5.7\times$, practical throughput improves by only $1.2\times$. This discrepancy arises from the GPU's inefficiency in fine-grained bit-level operations and merging redundant elements, resulting in exposed latency bottlenecks when identifying repetitive elements. After adding the dedicated CAM-based BRCR engine, the performance jumps by $2.88\times$. Similarly, directly applying bit-sparsity BSTC scheme and bit-prediction BGPP scheme yields only $1.44\times$ and $1.23\times$ gain, as the value-to-bit reorder cost and mismatched computation granularity, which lead to severe underutilization of GPU resources. By contrast, employing tailored engines can further bring $2.19\times$ and $1.48\times$ acceleration effects. Interestingly, although BSTC achieves smaller performance gains ($2.19\times$) than BRCR ($2.88\times$) on ASICs, it yields greater improvements ($1.44\times$ than $1.2\times$) on GPUs. This is primarily because BSTC significantly reduces memory access overhead, allowing GPUs, despite lacking dedicated encoding/decoding support, to still benefit from the optimization. A similar trend is observed with BGPP, which reduces memory access through token sparsification, thus enabling performance gains on GPUs as well.

\begin{figure}[t]
\centering
\includegraphics[width=\linewidth]{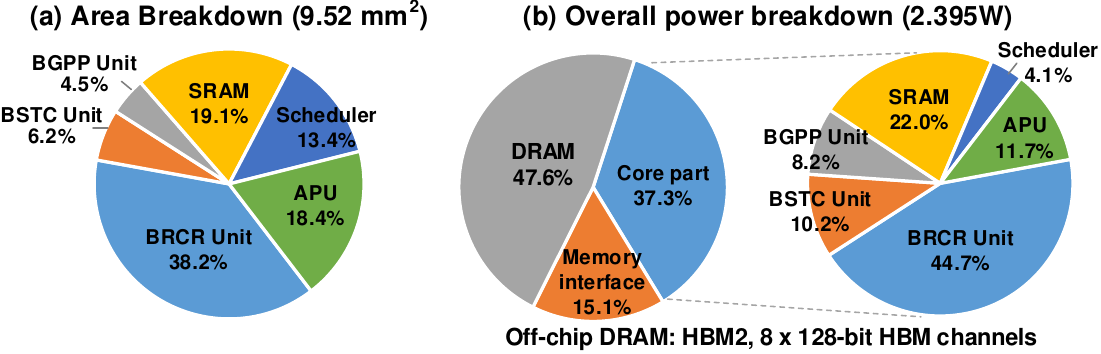}\vspace{-2mm}
\caption{Area/Power of MCBP at TSMC 28nm, 1GHz.}
\label{fig:Area_power}
\end{figure}

\emph{Area, Power and Energy:} Table \ref{tab:core_hardware_result} summarizes the hardware configuration of MCBP and Fig. \ref{fig:Area_power} shows its area and power breakdown. Here, we scale up the MCBP accelerator to contain 16 PE clusters to match the HBM I/O interface. The total power includes the core logic, memory interface \cite{leibowitz20104}, and external HBM. It has a total area of $9.52$ mm$^2$ and 2.395W power consumption. Benefted by the lightweight design of BSTC encoders and decoders, CODEC part accounts for merely $6.2\%$ and $10\%$ of area and core part power. Despite average $75.8\%$ reduction in IO traffic, DRAM power still accounts for approximately $48\%$ of total power consumption, due to the autoregressive nature of LLMs. Fig. \ref{fig:GPU_gain} (b) shows the overall energy-efficiency gain of MCBP over the A100 GPU. On average, MCBP standard/aggressive achieves $29.2\times$/$31.1\times$ greater efficiency than running dense benchmarks on GPU. Compared to naively running algorithm mechanism on GPU, MCBP standard/aggressive realizes $21.6\times$/$23.1\times$ gain. Fig. \ref{fig:Gain_breakdown} (b) also shows the efficiency gain breakdown. Software-hardware co-design BRCR, TSBC and BGPP bring $4.24\times$, $2.98\times$ and $2.44\times$ efficiency gain, respectively. 

\begin{table}[t]
\renewcommand{\arraystretch}{1.1}
\caption{Hardware Configurations of MCBP.}\vspace{-4mm}
\begin{center}
\footnotesize
\begin{tabular}{l|c}
\specialrule{0.12em}{0.5pt}{0.5pt}
\!\!\textbf{Main Modules} & \textbf{Parameters} \\
\hline
\!\!\!\multirow{1}{*}{\textbf{CAM-based BRCR Unit}} & 20 PE Clusters (160 PEs)   \\
\hline
\multirow{2}{*}{~~~~~\hspace{2mm}Processing Element (PE)}\!\!\!\!\!\!\! & One 512B CAM unit;~ 16 index converters \\
 & 16 Add merge units; 1 Reconstruction unit \\
\hline
\!\!\!\multirow{1}{*}{\textbf{BSTC CODEC Unit}} & $20\times4$ decoders; ~~~ $10\times4$ encoders  \\
\hline
\!\!\!\multirow{2}{*}{\textbf{Clock-gated BGPP Unit}} & $64$ 64-input AND-based Adder-trees    \\
& 4 Clock-gated Progressive Filters     \\
\hline
\!\!\!\textbf{Auxiliary Processing Unit}\!\! & \!\!\!1 Embedding unit;\,1 Special function unit;\,1 Quantizer\!\!\!\!\! \\
\hline
\!\!\!\multirow{2}{*}{\textbf{On chip Buffer}}  & \!\!\!$384$KB\,Token\,SRAM; $768$KB\,Weight\,SRAM\!\    \\
\!\! & \!\!$96$KB\,Temp\,SRAM\!\!    \\
\hline
\!\!\!\textbf{Main Memory} & \multicolumn{1}{c}{HBM2, 8$\times$128-bit HBM channels @2GHz, 8GB} \\
\specialrule{0.12em}{0.1pt}{0.1pt}
\end{tabular}
\end{center}
\label{tab:core_hardware_result}\vspace{-0mm}
\end{table}

\subsection{Comparison with SOTA Accelerators}
\begin{figure}[t]
\centering
\includegraphics[width=\linewidth]{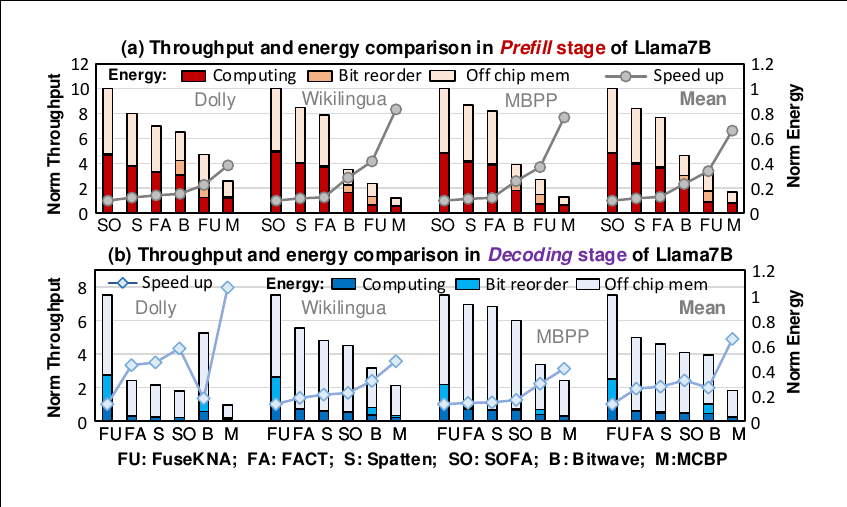}\vspace{-2mm}
\caption{Speedup on Llama7B (a) Prefill (b) Decoding.}
\label{fig:SOTA_acceleration_comparison}
\end{figure}

Fig. \ref{fig:SOTA_acceleration_comparison} compares the throughput and energy of various accelerators during prefill and decoding. Energy is broken down into compute, bit reordering, and off-chip memory. In the prefill stage (Fig.~\ref{fig:SOTA_acceleration_comparison}(a)), computation consistently accounts for over $30\%$ of total energy across all designs. Bit-reordering overhead is significant in FuseKNA ($30\%$) and BitWave ($18\%$) due to their value- or multi-bit compression schemes, which misalign with bit-serial processing. In contrast, MCBP limits this overhead to $3\%$ via a bit-slice–first encoding strategy. In terms of throughput, for long-sequence tasks like Dolly with high token-level sparsity, traditional Transformer accelerators like SOFA, Spatten, and FACT achieve notable speedup. In this case, MCBP offers a smaller advantage over Energon ($3.8\times$). However, in short-sequence tasks like MBPP, where token sparsity diminishes, bit-level redundancy becomes more exploitable. FuseKNA gains $3.7\times$ speedup via bit-repetition but suffers from high-latency serial matching. In contrast, MCBP achieves the best acceleration, with an average speedup of $6.2\times$.

In the \emph{decoding} stage (Fig.\ref{fig:SOTA_acceleration_comparison} (b)), speedup mainly comes from reduced memory access for weights and KV cache. For long-text tasks like Dolly, where the KV cache size exceeds the weight size, attention optimization in SOFA, Spatten, and FACT yields a $3.7\times$ speedup. Bitwave, optimizing only weights, achieves just $1.3\times$ speedup. As sequence length shortens (e.g., Wikilingua), the KV cache shrinks proportionally, leading to performance degradation in traditional Transformer accelerators. For code generation tasks (MBPP), Bitwave benefits from its weight-centric design. Across all workloads, MCBP achieves the highest performance, averaging $4.8\times$ speedup.

Table \ref{tab:hardware_compara} summarizes the specifications for SpAtten, FACT, and SOFA. They are the SOTA accelerators that exploit the attention mechanism to improve the energy efficiency of Transformer inference. Spatten and SOFA are both optimized for the attention. FACT focus on the whole model computation acceleration via top-$k$ token pruning. However, these works are all designed for prefill stage by finding redundancy of attention or linear computation, making them unsuitable for decoding stage in LLM. In addition, their optimizations all remain at value level, missing abundant opportunities at bit level. To the best of our knowledge, MCBP is so far the first work that uses bit-level strategies for LLM inference, reducing both memory and computation effort for both prefill and decoding stages of LLM. The energy efficiency of MCBP (with bit-level GEMM, weight $\&$ KV cache optimizations) is 22740 GOPS/W, which is $35\times$, $5.2\times$ and $3.2\times$ greater than the three counterparts, with different technology normalized to 28nm for fair comparison. The average energy efficiency is evaluated using the metric from each respective paper. However, SOFA experiences over a $4\times$ efficiency degradation when applied to autoregressive LLMs, as it is tailored for parallel processing of attention and fails to address the memory access bottleneck in LLMs. In contrast, MCBP achieves a $12.8\times$ efficiency gain over SOFA when processing LLMs.

\begin{table}[t]
\footnotesize
\renewcommand{\arraystretch}{1.1}
\caption{Summary and comparison with SOTA works.}\vspace{-4mm}
\begin{center}
\begin{tabular}{l|m{0.9cm}<{\centering}m{1.4cm}<{\centering}m{0.9cm}<{\centering}m{1.21cm}<{\centering}}
\specialrule{0.12em}{0.5pt}{0.8pt}
 & \!\!\!\textbf{SpAtten}\cite{wang2021spatten}  & \textbf{FACT}\cite{qin2023fact} & \textbf{SOFA}\cite{wang2024sofa} & \textbf{MCBP} \\
\hline 
\!\!\!\multirow{2}{*}{\textbf{Acceleration for}} & Prefill & Prefill & Prefill & \textbf{P $\&$ D}  \\
 & \!\!(attention) & \!\!\!\!(whole model)\!\!\!\! & \!\!(attention)\!\! & \!\!\!\!\textbf{whole model}\!\! \\
\!\!\!\textbf{Optimization level}$^{\ddagger}$ & Value C. & Value G.C.  & Value C. & \!\!\textbf{Bit G.W.C.}\!\!  \\
\!\!\!\textbf{Technology [nm]} & $40$ & $28$  & $28$ & $\mathbf{28}$  \\
\!\!\!\textbf{Area [mm$^2$]} & $1.55$ & $6.03$  & $4.29$ & $\mathbf{9.52}$ \\
\!\!\!\textbf{Throughput [GOPS]} & $360$ & $1153$  & $24,423$ & $\mathbf{54,463}$ \\
\!\!\!\textbf{Energy Effi. [GOPS/W]}\!\!\! & $382$ & $4388$  & $7183$ &$\mathbf{22,740}$\\
\specialrule{0.12em}{0.5pt}{0.5pt}
\end{tabular}\vspace{-1mm}
\begin{flushleft}
\scriptsize
\item $^{\ddagger}$ G: GEMM, W: Weight access. C: KV cache access. And optimizing at the value or bit-level.
\end{flushleft}
\end{center}
\label{tab:hardware_compara}\vspace{-0mm}
\end{table}

\section{Discussion}
\begin{figure}[t]
\centering
\includegraphics[width=\linewidth]{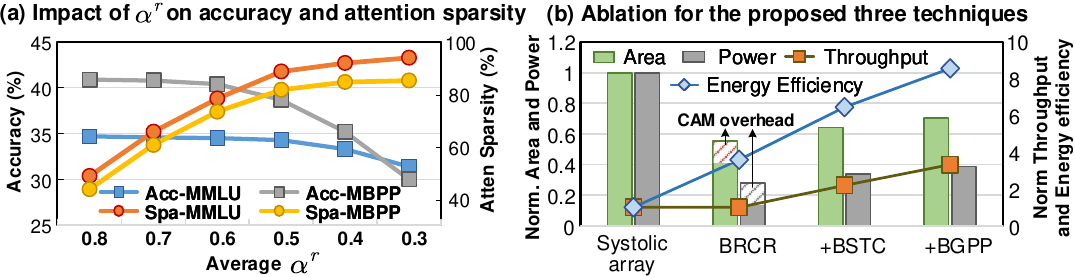}\vspace{-3mm}
\caption{(a) Evaluation of MCBP's optimization impact on inference accuracy. (b) Ablation study of the hardware overhead introduced by the three optimizations.}
\label{fig:disscusion_for_quantization}
\end{figure}

Among the three optimizations in MCBP, BRCR and BSTC are lossless, since they leverage intrinsic data redundancy and sparsity for acceleration. In contrast, BGPP introduces a hyperparameter $\alpha^r$ to select vital KVs (a.k.a., attention sparsity, \S\ref{subsec:KV Cache access optimization}), which may affect accuracy. Fig. \ref{fig:disscusion_for_quantization} (a) evaluate the impact of $\alpha^r$ on accuracy and attention sparsity using LLaMA-7B on two tasks: MMLU (reasoning) and MBPP (generation). Overall, a smaller $\alpha^r$ results in more aggressive pruning, which decreases model accuracy but increases sparsity. There are some key observations from Fig. \ref{fig:disscusion_for_quantization} (a): For generation tasks (MBPP), accuracy drops noticeably when $\alpha^r<0.6$. In contrast, for reasoning tasks (MMLU), the model is more tolerant to pruning, with performance degrading significantly only when $\alpha^r$$<$$0.5$. This may be because reasoning tasks rely on key tokens for inference, resulting in higher token redundancy. On the other hand, the sparsity gains begin to diminish when $\alpha^r$$<$$0.5$, this may be because overly aggressive pruning hurts some critical tokens. Therefore, to strike a well balance between accuracy and sparsity, we set $\alpha^r$ in the range of 0.5–0.6 in MCBP.

Fig. \ref{fig:disscusion_for_quantization} (b) presents an ablation exploration for BRCR, BSTC and BGPP, in terms of area and energy overhead, using a systolic array (SA) that provides the same throughput as the baseline. Although CAM adds $25\%$ area and $47\%$ power overhead to the BRCR unit, BRCR still reduces overall area and power by $45\%$ and $72\%$, respectively, while boosting energy efficiency by $3.6\times$. These gains stem from BRCR’s efficient use of bit-level redundancy to eliminate redundant computations (\S\ref{subsec:Bit_serial_Computation_Optimization}). Additionally, the integration of CAM enhances pipeline efficiency, contributing to overall performance gains. Building upon this, BSTC applies bit sparsity optimization, achieving a $2.2\times$ throughput gain with only $16\%$ area and $20\%$ energy overhead, driven by significantly reduced memory access. Finally, BGPP achieves a further $1.48\times$ throughput gain with just $9\%$ area and $13\%$ energy overhead, owing to the reduction in attention computation and associated memory access operations.

\begin{figure}[t]
\centering
\includegraphics[width=\linewidth]{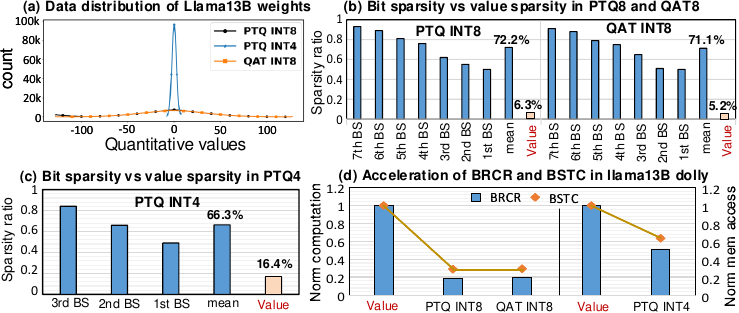}\vspace{-2mm}
\caption{The bit sparsity for diverse quantization scenarios.}
\label{fig:quantization}
\end{figure}

To explore bit-level sparsity across various quantization strategies, we profile Llama13B's weights under QAT INT8, PTQ INT8, and PTQ INT4, as shown in Fig. \ref{fig:quantization}(a). The weight distributions for QAT and PTQ INT8 are similar, likely due to LLMs’ fault tolerance enabling effective INT8 quantization. In contrast, PTQ INT4 exhibits a more concentrated distribution due to its lower bit width.

We compare bit and value sparsity across the three quantization strategies. The 7th BS denotes the highest bit-slice matrix (excluding the sign bit), while the 1st BS stands for the lowest. Fig. \ref{fig:quantization} (b) shows the average bit sparsity for PTQ and QAT INT8 is about $11\times$ higher than value sparsity. In contrast, PTQ INT4 notably increases value sparsity to $\sim16\%$ (Fig.\ref{fig:quantization}(c)), but bit sparsity remains higher at $66\%$ ($\sim 4\times$ higher). Fig. \ref{fig:quantization}(d) shows that BRCR reduces computation by $80\%$, $79.45\%$, and $51\%$ for PTQ INT8, QAT INT8, and PTQ INT4, respectively, while BSTC cuts memory accesses by $71\%$, $70.5\%$, and $41\%$. These results highlight MCBP’s broad effectiveness, driven by the greater prevalence of bit-level sparsity.

\begin{figure}[t]
\centering
\includegraphics[width=\linewidth]{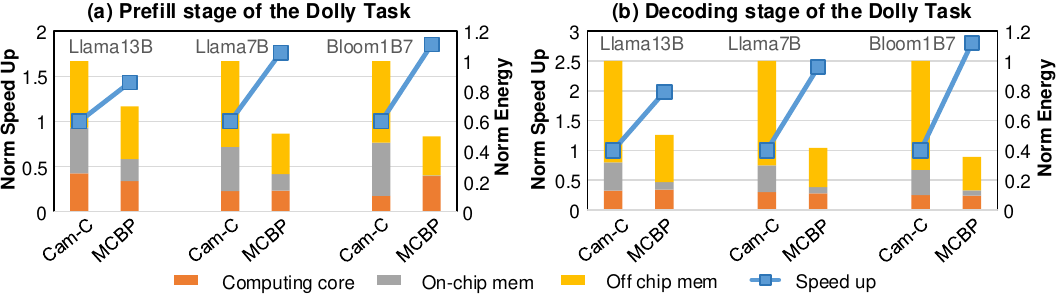}\vspace{-3mm}
\caption{Compared with SOTA INT4 accelerator.}
\label{fig:INT4_Accelerator}
\end{figure}

Cambricon-C (Cam-C) \cite{chen2024cambricon} is a SOTA INT4 accelerator that achieves computational acceleration by efficiently looking up all 256 product results (INT4×INT4, W4A4), avoiding explicit computation. We reproduce Cam-C with the same PE array area (4.65mm²) and 1248kB of on-chip SRAM as used in MCBP.
Considering that W4A4 quantization is too aggressive and typically results in a $4$–$6\%$ accuracy degradation \cite{hu2025quad} on modern LLMs, we adopt a more conservative W4A8 quantization for comparisons. Accordingly, we extend Cam-C to support W4A8, while retaining its core optimization technique—Quarter Square Multiplication. Using the QLLM framework \cite{liu2023qllm}, we quantize BloomB7, Llama7B/13B to W4A8, ensuring an accuracy loss of less than $1\%$ compared to FP16. We evaluate the performance of Cam-C and MCBP on Dolly datasset of the three models and draw the following three key observations: 

First, Cam-C suffers from significant look-up overhead. This is because Cam-C relies on the look-up of all possible results. When the activation is extended to INT8, the cost of look-up increases dramatically, limiting Cam-C's acceleration. This limitation is particularly evident with small models, e.g. Bloom1B7, where value-level redundancy cannot be guaranteed as their small hidden sizes. As shown in Fig. \ref{fig:INT4_Accelerator} (a), compared to Cam-C, MCBP achieves $1.5\times$ speedup and $33\%$ energy savings on LLaMA-13B, and $1.8\times$ speedup with $50\%$ lower energy consumption on Bloom-1B7. This benefits from MCBP maximizes redundancy utilization unexploited at the \textbf{bit level} instead of value level.
Second, Cam-C fails to leverage the inherent bit sparsity of INT4 and attention sparsity for memory optimization, leading to poor performance during decoding stage. In contrast, MCBP utilizes BSTC to exploit the sparsity of INT4 and BGPP for KV cache traffic reduction, reducing memory access and achieving an average $2.4\times$ speedup, as depicted in Fig. \ref{fig:INT4_Accelerator} (b).

Overall, MCBP demonstrates an evident performance advantage over the SOTA INT4 accelerator, thanks to its comprehensive optimization of the LLM inference bottleneck at the bit level.

\section{Related Works}\label{sec:related_work}
\textbf{Transformer accelerator}. Numerous works \cite{li2020ftrans, ham20203, ham2021elsa, qu2022dota, fan2023taskfusion, yang2022dtatrans, hong2022dfx, liu2024hsconn, wang2024sofa, wang2022energy, fang2022algorithm, yazdanbakhsh2022sparse, li2022accelerating, lu2021sanger, wang2021spatten, zhou2022energon, qin2023fact, shen2022salo, fan2022adaptable, qin2024mecla, bai2024swat} have been proposed to improve efficiency of Transformer-based LLMs. Given the quadratic complexity of attention in long sequences \cite{vaswani2017attention}, many focus on accelerating attention via static sliding windows \cite{you2023vitcod, li2020ftrans, shen2022salo, bai2024swat, fan2022adaptable} or dynamic top-$k$ prediction \cite{ham20203, ham2021elsa, qu2022dota, liu2024hsconn, wang2024sofa, yazdanbakhsh2022sparse, li2022accelerating, lu2021sanger}. Other works extend token sparsity to linear layers \cite{yang2022dtatrans, wang2021spatten, li2020ftrans, qin2023fact, qin2024mecla, wang2022energy, fang2022algorithm, zhou2022energon, fan2022adaptable}. However, due to the autoregressive nature of LLMs, weight and KV cache memory access dominate latency during the decoding stage—an aspect largely overlooked by prior designs. In contrast, MCBP addresses these bottleneck holistically. In the \textit{prefill} stage, MCBP optimizes computation with bit-slice (BS) repetition, while in the \textit{decoding} stage, it minimizes memory access through bit sparsity and bit-grained early termination.

\textbf{Value sparsity accelerator}. Numerous accelerators \cite{liu202316, tambe2021edgebert, fang2022algorithm, fang2022efficient, tuli2023acceltran, yoo2023tf, you2023vitcod, deng2021gospa, kang2021ganpu, liu2022s2ta, mahmoud2020tensordash, hanson2022cascading, li2021escalate, shin2022griffin, wu2023highlight, gudaparthi2022candles, lew2022anticipating, yang2020procrustes, han2016eie, qin2020sigma, zhang2016cambricon, moons2016energy, han2015deep, moons201714} exploit value sparsity to improve NNs performance. EIE \cite{han2016eie} utilizes dynamic input and static weight sparsity to accelerate CNNs and RNNs, while S2TA \cite{liu2022s2ta} leverages structured sparsity in weights and activations to accelerate CNNs. Recent efforts \cite{liu202316, tambe2021edgebert, fang2022algorithm, fang2022efficient, tuli2023acceltran, yoo2023tf} have extended this idea to LLMs, e.g., EdgeBERT \cite{tambe2021edgebert} uses masks to skip zeros in weights to reduce unnecessary computation. However, value sparsity in LLMs is highly limited (Llama13B 6.3\%). In contrast, MCBP uses extremely fine-grained bit sparsity, which is average $10.1\times$ higher than value sparsity, bit repetition, and bit prediction to remove redundant memory access.

\textbf{Bit-serial computing accelerators}. Prior works \cite{kam2024panacea,im2023sibia,han2024bitnn,yang2021fusekna,judd2016stripes,albericio2017bit,delmas2019bit,lo2023bit,sharify2019laconic,lee2018unpu,delmas2017dynamic,gondimalla2019sparten,li2022ristretto,parashar2017scnn,yang2023isosceles,lu2021distilling,kam2024panacea,li2025lut,im2023sibia} accelerates neural networks by exploiting bit-level sparsity within individual BS vector \cite{albericio2017bit,delmas2019bit,lu2021distilling,sharify2019laconic,yang2021fusekna} or dynamically reducing bit-width \cite{gondimalla2019sparten,li2022ristretto,parashar2017scnn,yang2023isosceles}. However, such techniques fall short for LLMs, which are both memory- and computation-intensive. Besides, they often target irregular activation sparsity \cite{yang2021fusekna,delmas2019bit,sharify2019laconic}, which doesn’t address LLM bottlenecks like weight and KV cache access. In contrast, MCBP eliminates redundancy across BS vectors and uses BS sparsity and fine-grained bit prediction to reduce weight and KV cache memory accesses.

\section{Conclusion}
We propose MCBP, a software-hardware co-design to accelerate the computation, weight access and KV cache access for LLM inference. Utilizing the bit-level repetition, sparsity and reduced prediction traffic, MCBP achieves $31.1\times$, $35\times$, $5.2\times$ and $3.2\times$ energy saving than A100 GPU, SOTA accelerators SpAtten, FACT and SOFA.

\begin{acks}
This work was supported in part by the  National Science and Technology Major Project under Grant 2022ZD0115200; the NSFC under Grant 62125403, and Grant 92164301; Beijing S$\&$T Project Z221100007722023; in part by the project funding for the 2022 Special Project on Industrial Foundation Reconstruction and High Quality Development of Manufacturing Industry CEIEC-2022-ZM02-0245; in part by the Beijing National Research Center for Information Science and Technology; and in part by the Beijing Advanced Innovation Center for Integrated Circuits. 
\end{acks}


\bibliographystyle{ACM-Reference-Format}
\bibliography{MCBP}

\end{document}